\documentclass[12pt,graphicx]{article}
\usepackage{amssymb}

\usepackage{amsmath}
\usepackage[dvips]{graphics}
\usepackage[dvips]{graphicx}

\newtheorem{lemma}{Lemma}

\newenvironment{proof}[1][Proof]{\textbf{#1.} }{\ \rule{0.5em}{0.5em}}

\input{tcilatex}

\numberwithin{equation}{section}
\numberwithin{figure}{section}

\newcommand{\benumerate}{\begin{enumerate}}
\newcommand{\eenumerate}{\end{enumerate}}

\newcommand{\bitemize}{\begin{itemize}}

\newcommand{\eitemize}{\end{itemize}}
\newcommand{\der}[2]{\frac{\partial #1}{\partial #2}}

\newcommand{\paper}[6]{#1, #2, {\em #3}, {\bf #4}, #5,  (#6)}
\newcommand{\paperr}[5]{#1, {\em #2} {\bf #3}, #4, (#5)}

\newcommand{\book}[4]{#1, {\em #2}, {#3}, (#4)}

\begin{document}

\date{}

\title{Integrable equations in nonlinear geometrical optics.}

\author{Boris Konopelchenko\footnote{Supported in part by the COFIN
PRIN ``SINTESI'' 2002.}~\footnote{e-mail: konopel@le.infn.it}~ and
Antonio Moro$^{\ast}$\footnote{e-mail: antonio.moro@le.infn.it} \\
{\em Dipartimento di Fisica dell'Universit\`{a} di Lecce} \\
{\em and INFN, Sezione di Lecce, I-73100 Lecce, Italy}}
\maketitle

\begin{abstract}
Geometrical optics limit of the Maxwell equations for nonlinear media
with the Cole-Cole dependence of dielectric function and magnetic
  permeability on the frequency is considered. It is shown that for
media with slow variation along one axis such a limit gives rise to
the dispersionless Veselov-Novikov equation for the refractive
index. It is demonstrated that the Veselov-Novikov hierarchy is
amenable to the quasiclassical $\bar{\partial}$-dressing method. Under
more specific requirements for the media, one gets the dispersionless
Kadomtsev-Petviashvili equation. Geometrical optics interpretation of
some solutions of the above equations is discussed. 

PACS numbers: 02.30.Ik, 42.15.Dp  \\
Key words: Nonlinear Optics, Integrable Systems.
\end{abstract}

\section{Introduction.}
\label{sec_intro}
Nonlinear optics provides us with various, very interesting nonlinear
phenomena (see e.g. \cite{Boyd,Fokas}). At the same time, it is the source of
several nonlinear differential equations, modelling such phenomena,
with certain very special properties. The nonlinear Schr\"odinger (NLS)
equation, which describes self-modulation and self-focusing of
electromagnetic waves in nonlinear media \cite{Boyd,Fokas} is, perhaps, the
most known one. The NLS equation is integrable by the inverse
scattering transform method and has a number of remarkable properties
(multisoliton solutions, infinite set of conserved quantities and so
on).
Nowadays there are a number of the, so-called, integrable nonlinear differential equations which have similar properties \cite{Zakharov,Segur}. A
subclass of such equations, referred usually as dispersionless integrable
equations, has attracted a particular interest in the last years (see e.g. \cite{LaxLev}-\cite{Dubrovin2}).

In the present paper we will study integrable structures which arise
within the geometrical optics of nonlinear media. We will consider
nonlinear media which are characterized by the Cole-Cole dependence of
dielectric function and magnetic permeability on the frequency and
slow variation of all quantities along one axis. We will show that the
propagation of monochromatic electromagnetic waves of high frequency
$\omega$ in such media is governed by the dispersionless
Veselov-Novikov (dVN) equation. Namely, we will demonstrate that the
Maxwell equation in such a situation in the limit $\omega \rightarrow
\infty$ gives rise to the standard plane eikonal equation accompanied
by the equation which describes its deformation in the orthogonal
direction.
The compatibility of these two equations is equivalent to dVN equation
for the refractive index. This dVN deformations of the plane eikonal
equation preserve, in particular, the total ``plane'' squared
refractive index $\int\int n^{2} dx dy$. We will show also that under
more specific conditions, the Maxwell equation is reduced to the
dispersionless Kadomtsev-Petviashvili (dKP) equation.

In this paper, we will demonstrate that the plane eikonal equation as well as the dVN
equation and whole dVN hierarchy, both for complex and real-valued
refractive indices are treatable by the quasiclassical
$\bar{\partial}$-dressing method recently developed in \cite{Konopelchenko2,Konopelchenko3,Konopelchenko4}.
The characterization conditions for the $\bar{\partial}$-data,
symmetry constraints for the dVN equation and associated hodograph
type solutions are also studied. 
Geometrical optics interpretation of some solutions of dKP  and dVN
equations is discussed too.

The paper is organized as follows. In section 2 we discuss general
properties of nonlinear media under consideration. In section 3 the
dVN equation is derived as the geometrical optics limit of the Maxwell
equations. The particular case of the dKP equation is discussed in
section 4. The quasiclassical $\bar{\partial}$-dressing method is
applied to the plane eikonal equation in the section 5. The dVN
equation is treated in the next section 6. Characterization conditions
for $\bar{\partial}$-data are considered in section 7. Symmetry
constraints for dVN equation and associated hodograph type solutions
are discussed in section 8. At the last section 9 we present several
exact and numerical solutions both for dVN and dKP equations and
corresponding wavefronts. 

Some results of this paper have been announced in the letter \cite{KonopMoro}.    

\section{Maxwell equations in nonlinear media.}
\label{sec_maxwell}
We begin with the Maxwell equations in the absence of sources (we
put the velocity of the light in the vacuum $c=1$)
\begin{eqnarray}
\label{maxwell}
&&\nabla \wedge {\bf H} - {\bf \dot{D}}=0~~~~~~\nabla \cdot {\bf D} = 0 \nonumber \\
&&\nabla \wedge {\bf E} + {\bf \dot{B}}=0 ~~~~~~~\nabla \cdot {\bf B} = 0
\end{eqnarray}
and the material equations of a medium
\begin{equation}
\label{materials}
{\bf D} = \varepsilon {\bf E},~~~~~~{\bf B} = \mu {\bf H}.
\end{equation}
Here and below $\nabla = \left(\partial_{x},\partial_{y},\partial_{z}
\right)$, while $\cdot$ and $\wedge$ denote the scalar and vector
product respectively.   

We will study a propagation of electromagnetic waves of the fixed,
high frequency $\omega$, i.e. we will look for the solutions of the
Maxwell equations of the form \cite{Born}
\begin{eqnarray}
\label{full_fields}
&&{\bf E}\left(x,y,z,t\right) = {\bf E_{0}}\left(x,y,z\right)
e^{-i\omega t} = {\bf \tilde{E}_{0}}\left(x,y,z\right)
e^{-i\omega t + i \omega S\left(x,y,z\right)} \nonumber\\
&&{\bf H}\left(x,y,z,t\right) = {\bf H_{0}}\left(x,y,z\right)e^{-i\omega t} = {\bf \tilde{H}_{0}}\left(x,y,z\right) 
e^{-i\omega t + i \omega S\left(x,y,z\right)}
\end{eqnarray} 
where ${\bf E_{0}}$, ${\bf H_{0}}$ and the phase $S\left(x,y,z
\right)$ are certain functions and ${\bf \tilde{E}_{0}}
=\sum_{n=0}^{\infty} \omega^{-n} {\bf E_{0}}_{,n}(x,y,z)$, ${\bf \tilde{H}_{0}}
=\sum_{n=0}^{\infty} \omega^{-n} {\bf H_{0}}_{,n}(x,y,z)$.

Maxwell equations~(\ref{maxwell}-\ref{materials}) imply the following
stationary second order equations for ${\bf E_{0}}$ and ${\bf
  H_{0}}$ \cite{Born} 
\begin{eqnarray}
\label{wave}
\nabla^{2}{\bf E_{0}} + \omega^{2} \mu \varepsilon {\bf E_{0}} + \left (\nabla
\log~\mu \right) \wedge \left(\nabla \wedge {\bf E_{0}} \right ) +
\nabla \left ({\bf E_{0}} \cdot \nabla \log~\varepsilon \right) &=& 0\nonumber \\
\nabla^{2}{\bf H_{0}} + \omega^{2} \mu \varepsilon {\bf H_{0}} + \left (\nabla
\log~\varepsilon \right) \wedge \left(\nabla \wedge {\bf H_{0}} \right ) +
\nabla \left ({\bf H_{0}} \cdot \nabla \log~\mu \right) &=& 0.
\end{eqnarray}
We assume that the medium is characterized by the Cole-Cole
dependence \cite{ColeCole} of the dielectric function $\varepsilon$ and the
magnetic permeability $\mu$, namely, that
\begin{equation}
\label{coledip1}
\varepsilon = \varepsilon_{0} + \frac{\tilde{\varepsilon}}{1+\left (i
  \omega \tau_{0} \right)^{2\nu}},~~~~~~0< \nu < \frac{1}{2}
\end{equation}
and 
\begin{equation}
\label{coledip2}
\mu = \mu_{0} + \frac{\tilde{\mu}}{1+\left (i
  \omega \tau_{0} \right)^{2\nu}},~~~~~~0< \nu < \frac{1}{2}
\end{equation}
where $\varepsilon_{0}$, $\mu_{0}$ depend on
the coordinates $x,y,z$, while  $\tilde{\varepsilon}$ and
$\tilde{\mu}$ depend both on the coordinates and the moduli of the fields and
their derivatives.
In the high frequency limit $\omega \rightarrow \infty$ we have
\begin{eqnarray}
\label{high_Cole}
\varepsilon = \varepsilon_{0}(x,y,z) + \frac{\tilde{\varepsilon_{1}}}{\omega^{2
\nu}} \nonumber \\
\mu = \mu_{0}(x,y,z) + \frac{\tilde{\mu_{1}}}{\omega^{2
\nu}}
\end{eqnarray}
where $\tilde{\varepsilon_{1}} = (i \tau_{0})^{-2 \nu} \tilde{\varepsilon}$ and  $\tilde{\mu_{1}} = (i \tau_{0})^{-2 \nu} \tilde{\mu}$.  
In some textbooks (see e.g. \cite{Landau,Jackson}) it is
argued that at $\omega \rightarrow \infty$ the dielectric function has
the following asymptotic behaviour
\begin{equation}
\label{landaudip}
\varepsilon(\omega) \rightarrow 1 +
\frac{{\sl const}}{\omega^{2}},~~~~\omega \rightarrow \infty.
\end{equation}
Experimental results of the Cole and Cole \cite{ColeCole} have
demonstrated that the theoretical models behind the
formula~(\ref{landaudip}) are not always correct. The fact that the
parameter $\nu$ in the Cole-Cole dependence~(\ref{coledip1})
and~(\ref{coledip2}) belongs to the interval $\left(0,\frac{1}{2}\right)$
is of crucial importance for our study. 

At limit $\omega \rightarrow \infty$,  $\varepsilon$ and $\mu$ may depend
on the derivatives $S_{x}, S_{y}, S_{z}$. Different mechanism may be
responsible for such a dependence. One is provided by a non-local
spatial dependence of ${\bf D}$ on ${\bf E}$.
Indeed, let us assume that a monochromatic wave of frequency $\omega$ of the
form~(\ref{full_fields}) propagating in a medium have a spatial
non-local dependence for the displacement $\bf{D}({\bf x}, t)$ on $\bf{E}({\bf x}, t)$ of the 
standard form
\begin{equation}
\label{nonlocal_displacement}
{\bf D_{0}}({\bf x},\omega) = \int\int\int_{\mathbb{R}^{3}} 
g({\bf x'-x},\omega) {\bf E_{0}}({\bf x'-x},\omega) d{\bf x'}.
\end{equation}
In the particular case where the distribution $g({\bf x},\omega)$ is
proportional to a $\delta$-Dirac function
\begin{equation}
g({\bf x},\omega) = \varepsilon({\bf x},\omega) \delta({\bf x'-x})
\end{equation}
one obtains the usual material equation~(\ref{materials}).
Here, we consider non-local effects involving a linear superposition
of derivatives $\delta$-functions in the following way
\begin{equation}
\label{distribution} 
g({\bf x},\omega) = \sum_{l,m,n=0}^{\infty} g_{ l,m,n}({\bf x},\omega)
\delta^{(l,m,n)}({\bf x'-x }).
\end{equation}
The distributions $\delta^{(l,m,n)}$ are defined standardly as
\begin{equation}
\int\int\int_{\mathbb{R}^{3}} f({\bf x'}) \delta^{(l,m,n)}({\bf
  x'-x}) d{\bf x'} = (-1)^{l+m+n} \frac{\partial^{l+m+n} f({\bf x})}{\partial
  x_{1}^{l} \partial x_{2}^{m} \partial x_{3}^{n}},
\end{equation}
for a function $f({\bf x})$ defined in $\mathbb{R}^{3}$ and ${\bf x} =
(x_{1},x_{2},x_{3})$. With the use of the~(\ref{distribution})
the formula~(\ref{nonlocal_displacement}) for the $jth$ component of the field,
for $j=1,2,3$ can be written as
\begin{equation}
D_{0j}({\bf x},\omega) = \varepsilon_{j} E_{0j}({\bf x},\omega) 
\end{equation}
where the factors $\varepsilon_{j}$ are given by
\begin{equation}
\label{tensor}
\varepsilon_{j}({\bf x},\omega) = \sum_{l,m,n=0}^{\infty}(-1)^{l+m+n}
\frac{g_{l,m,n}({\bf x},\omega)}{E_{0j}({\bf x},\omega)}
\frac{\partial^{l+m+n} E_{0j}({\bf x},\omega)}{\partial x_{1}^{l} \partial
  x_{2}^{m} \partial x_{3}^{n}}. 
\end{equation}
Assuming that in the limit $\omega \rightarrow \infty$ $g_{000}$ does
not depend on the frequency, while
\begin{equation}
g_{l,m,n}({\bf x},\omega) \longrightarrow \frac{1}{\omega^{2 \nu +
    l+m+n}}\tilde{g}_{l,m,n}({\bf x});~~~~~~\omega \rightarrow \infty,
\end{equation}
one gets the formula~(\ref{high_Cole}) from~(\ref{tensor}), where
$\tilde{\varepsilon}$ and $\tilde{\mu}$ depend on $S_{x}$, $S_{y}$ and $S_{z}$.

We note that a number of dielectrics has the Cole-Cole dependence of
dielectric function $\varepsilon$ on $\omega$ \cite{ColeCole,ColeRecent1,ColeRecent2}, while for most
of studied media the magnetic permeability $\mu$ does not depend on
$\omega$.
In this paper, however, for the sake of completeness we will consider
the general case~(\ref{high_Cole}). Note also that at $\omega \rightarrow \infty$ certain typical effects on
nonlinearity, like a second-harmonic generation, become irrelevant.

In addition to the Cole-Cole property we assume that the medium is
anisotropic one, that is all quantities (${\bf E}$, ${\bf H}$, $\varepsilon$,
$\mu$) vary slowly along the axis $z$, such that, formally, $f_{z}
= \omega^{-\nu} f_{\xi}$, (here and below $f_{z} = \der{f}{z}$
etc.) and $\xi$ is a ``slow'' variable
defined by $z = \omega^{\nu} \xi$. More precisely, we assume that at
$\omega \rightarrow \infty$
\begin{equation}
\label{nuexpansion}
f(x,y,\omega^{\nu}\xi) = f_{0}(x,y,\xi)+\omega^{-\nu} f_{1}(x,y,\xi) +
\omega^{-2 \nu} f_{2}(x,y,\xi)+ \dots.
\end{equation}

We would like to note that the Cole-Cole property~(\ref{high_Cole})
is quite generic and has been widely discussed, while the particular
type of slow variation along the axis $z$ given
by~(\ref{nuexpansion}), seems, did not attract attention before.

\section{dVN equation as the geometrical optics limit of the Maxwell equations.}\label{sec_geometrical}  
Now we will analyze the squared Maxwell equations~(\ref{wave}) for the
media described in section~\ref{sec_maxwell}, i.e. for solutions of
the form~(\ref{full_fields}), where at $\omega \rightarrow \infty$
\begin{eqnarray}
\label{Snuexpansion}
S(x,y,\omega^{\nu}\xi) = S(x,y,\xi)+\omega^{-\nu} S_{1}(x,y,\xi) +
\omega^{-2 \nu} S_{2}(x,y,\xi)+ \dots, \\
\label{epsexpansion}
\varepsilon(x,y,\omega^{-\nu}\xi) = \varepsilon_{0}(x,y,\xi) + \omega^{-\nu}
\varepsilon_{1}(x,y,\xi) + \omega^{-2\nu} \varepsilon_{2}(x,y,\xi)+ \dots, \\
\label{muexpansion}
\mu(x,y,\omega^{-\nu}\xi) = \mu_{0}(x,y,\xi) + \omega^{-\nu}
\mu_{1}(x,y,\xi) + \omega^{-2\nu} \mu_{2}(x,y,\xi)+ \dots
\end{eqnarray}
Taking into account these expansions, one gets in the leading
$\omega^{2}$ order the plane eikonal equation
\begin{equation}
\label{eikonal}
S_{x}^{2}+S_{y}^{2} = n^{2}
\end{equation}
where  $n^{2}=\varepsilon_{0} \mu_{0}$.
One, obviously, obtains this plane eikonal equation if there is no
dependence on $z$ at all.

In the next $\omega^{2-\nu}$ and $\omega^{2-2\nu}$ orders, one gets
from~(\ref{wave}) the following equations
\begin{eqnarray}
\label{secondorder}
&&2 S_{x}S_{1x} + 2 S_{y}S_{1y} = \varepsilon_{1} \mu_{0} + \varepsilon_{0}
\mu_{1} \\
\label{thirdorder}
&&S_{\xi}^{2}= - \left(S_{1x}^{2} + S_{1y}^{2} + 2 S_{x}S_{2x} + 2
S_{y} S_{2y}\right) + \varepsilon_{0} \mu_{2} + \varepsilon_{1} \mu_{1} + \varepsilon_{2} \mu_{0}.
\end{eqnarray}
As we noted in the previous section, $\varepsilon$ and $\mu$ depend, in
general, on $S_{x}$,$S_{y}$ and $S_{z}$. Having in
mind~(\ref{Snuexpansion}-\ref{muexpansion}), one concludes that in the order
$\omega^{2-\nu}$, $\varepsilon_{1}$, $\mu_{1}$
might depend on the coordinates and only on $S_{x}$ and $S_{y}$. So
equation~(\ref{secondorder}) defines $S_{1}$ in terms of $S_{x}$ and
$S_{y}$. The function $S_{2}$ in equation~(\ref{thirdorder}) is an
undetermined function which in turn might depend on $S_{x}$,$S_{y}$
only.
Thus, equation~(\ref{thirdorder}) assumes the form
\begin{equation}
\label{csideformation}
S_{\xi} = \varphi\left(x,y,\xi;S_{x},S_{y} \right)
\end{equation}
where $\varphi$ is certain function collecting the contributions of
$S_{1}$, $S_{2}$ and $\varepsilon_{j}$, $\mu_{j}$, $j=0,1,2$. 
Let us note that the dependence of the function $\varphi$ on $S$ is not
admissible. Indeed, for the solutions of the form~(\ref{full_fields}) the time-translations symmetry
$t \rightarrow t+const$
of the Maxwell equations is equivalent to the phase displacement $S
\rightarrow S+const $. In order to preserve this symmetry in
equation~(\ref{csideformation}), $\varphi$ should depend only on
derivatives of the phase function $S$.

Introducing the complex variables $z = x+iy$, $\bar{z} = x-iy$, one
rewrites equations~(\ref{eikonal}) and~(\ref{csideformation}) as follows 
\begin{eqnarray}
\label{eikonal_complex}
&& S_{z} S_{\bar{z}} = u(z,\bar{z},\xi) \\
\label{z_direction_complex}
&& S_{\xi} = \varphi\left (z,\bar{z},\xi;S_{z},S_{\bar{z}} \right)
\end{eqnarray} 
where $u = 4 n^{2}$. \\
The condition of compatibility of equations~(\ref{eikonal}) and~(\ref{csideformation}) (or
~(\ref{eikonal_complex}), (\ref{z_direction_complex})) imposes
constraints on the possible forms of the function $\varphi$, namely
\begin{equation}
\label{compatibility}
S_{\bar{z}} \varphi_{z} + S_{z}
\varphi_{\bar{z}} + u_{z} \varphi' + u_{\bar{z}}
\varphi'' = u_{\xi},
\end{equation}
where 
\begin{equation}
\varphi' =
\der{\varphi}{S_{z}}\left(z,\bar{z};S_{z},S_{\bar{z}}
\right),~~~~~~\varphi'' =
\der{\varphi}{S_{\bar{z}}}\left(z,\bar{z};S_{z},S_{\bar{z}}
\right).
\end{equation}
Here we restrict ourself by functions $\varphi$ which are polynomial
in $S_{z}$, $S_{\bar{z}}$ and compatible with real-valuedness of $S$ and $u$.
For the simplest choice $\varphi = \alpha_{0}(z,\bar{z},\xi)$,
equation~(\ref{compatibility}) obviously gives $\alpha_{0}=const$,
i.e. $u_{\xi}=0, S = \alpha_{0} \xi +
\tilde{S}(z,\bar{z})$. For the linear function $\varphi =
\alpha S_{z} + \bar{\alpha} S_{\bar{z}}+\beta + \bar{\beta}$, one gets
$\alpha = \alpha(z)$, $\beta=const$, and 
\begin{equation}
\label{linear}
u_{\xi} = \left(\alpha u \right)_{z} + \left(\bar{\alpha} u
\right)_{\bar{z}}.
\end{equation}
 For the quadratic
\begin{equation}
\label{quadratic_case}
\varphi = \alpha S_{z}^{2} + \bar{\alpha}
S_{\bar{z}}^{2}+\beta S_{z}+ \bar{\beta} S_{\bar{z}}+
\gamma + \bar{\gamma},
\end{equation} 
equations~(\ref{compatibility})
and~(\ref{eikonal_complex}) imply $\alpha = 0$, $\beta = \beta(z)$ and
$\gamma = const$, i.e. one gets the previous linear case~(\ref{linear}).

The cubic
\begin{equation}
\label{cubic_case}
\varphi = \alpha S_{z}^{3} + \bar{\alpha} S_{\bar{z}}^{3}+ \beta S_{z}^{2}+\bar{\beta} S_{\bar{z}}^{2}+
\gamma S_{z} + \bar{\gamma} S_{\bar{z}} + \delta + \bar{\delta}
\end{equation}
obeys equations~(\ref{compatibility}) and~(\ref{eikonal_complex}) if
\begin{eqnarray}
&&\alpha = \alpha(z),~~~~~~\beta = 0, \nonumber \\
&&\gamma_{\bar{z}} = - \alpha_{z} u - 3 \alpha u_{z},
\nonumber \\
&&\delta = const,
\end{eqnarray}
and one has the equation 
\begin{equation}
\label{u_cubic}
u_{\xi} = \left(\gamma u\right)_{z} + \left(\bar{\gamma} u
\right)_{\bar{z}}. 
\end{equation}
In the particular case $\alpha=1$ and,
consequently $\gamma_{\bar{z}} = -3 u_{z}$, equation~(\ref{u_cubic}) is nothing
but the dispersionless Veselov-Novikov (dVN) equation introduced
in \cite{Krichever,Konopelchenko4}.

In a similar manner one can construct higher order nonlinear deformations of
wavefronts along $\xi$-direction which
correspond to higher degree polynomials $\varphi$.
These higher degree
cases apparently become physically relevant for the phenomena with
large values of $S_{x}$ and $S_{y}$.
Thus, if we formally admit all possible degrees of $S_{z}$ and
$S_{\bar{z}}$ in the right hand side of
equation~(\ref{z_direction_complex}), then one has an infinite family
of nonlinear equations, which may govern the $\xi$-variations of the
wavefronts and ``refractive index'' $u$. Since 
equation~(\ref{z_direction_complex}) should respects the
symmetry $S \rightarrow - S$ of the eikonal
equation~(\ref{eikonal_complex}), one readily concludes that only polynomials $\varphi$ of
the form $\sum_{m=1}^{n} u_{m} S_{z}^{2m-1} + \sum_{m=1}^{n}
\bar{u}_{m} S_{\bar{z}}^{2m-1}$ , are
admissible (the constant terms which have appeared in the
cases~(\ref{linear}), (\ref{u_cubic}) discussed above are, in fact,
irrelevant). Hence, these polynomial deformations are given by
\begin{equation}
\label{higherpolynomials}
S_{\xi} = \sum_{m=1}^{n} \left(u_{m} S_{z}^{2m-1} + \bar{u}_{m} S_{\bar{z}}^{2m-1}\right), ~~~~~n=1,2,3,\dots
\end{equation}
where $u_{m}$ are certain functions on $u$.

In the case $u_{n}=1$ one gets the dVN equation mentioned above ($n=2$) and
the so-called dVN hierarchy of nonlinear equations. The dVN equation
has been introduced in \cite{Krichever,Konopelchenko4} as the dispersionless limit of the VN
equation, which is the $2+1$-dimensional integrable
generalization of the famous Korteweg-de-Vries (KdV)
equation (see e.g. \cite{Zakharov,Segur,Konopelchenko}).

The dVN equation~(\ref{u_cubic}) and the whole dVN hierarchy have an
infinite set of integrals of motion \cite{KonopMoro}. The simplest of them is given by
$C_{1} = \int \int u dx dy$. Indeed, in virtue of
equation~(\ref{u_cubic})
\begin{eqnarray}
\label{int_motion}
C_{1\xi} = \int\int_{G} u_{\xi} dx dy &=&  \int\int_{G} \left(\left(\gamma u \right)_{z} + \left(\bar{\gamma}
u\right)_{\bar{z}} \right) dx dy = \\
&=& \int_{\partial G} \left(\frac{1}{2 i}~\bar{\gamma} u ~dz + c.c. \right)  
\end{eqnarray}
where $G$ is a domain in $\mathbb{C}$ and $\partial G$ is the boundary
of $G$. For $G=\mathbb{C}$ and solutions of dVN equation
such that $\gamma u \rightarrow 0$ at $|z| \rightarrow \infty$
\begin{equation}
C_{1\xi} = \int\int_{\mathbb{C}}u_{\xi} dx dy = 0.
\end{equation}
Thus the dVN hierarchy represents itself the
class of deformations of the plane eikonal
equation~(\ref{eikonal_complex}), which preserve the total ``plane''
squared refractive index  $\int \int n^{2}(x,y,\xi) dx dy$. The
physical meaning of higher conserved quantities is not that clear yet.

Note that for the finite domain $G$, $C_{1}$ is proportional to the
Dirichlet integral over domain $G$, $C_{1} = 4 \int\int
\left(S_{x}^{2}+S_{y}^{2} \right) dx dy$. The
formula~(\ref{int_motion}) gives us the variations of the Dirichlet
integral due to the dVN deformations.

\section{Integrable deformations of the quasiplane wavefronts via dKP
  equation.}  
Let us consider a more specific situation in which the
propagation of electromagnetic waves in the media discussed above
exhibits also a slow variation along the axis $y$, namely $\partial_{y}
= \epsilon \partial_{\eta}$, where $\eta$ is a slow variable defined
by $y = \frac{\eta}{\epsilon}$ and $\epsilon$ is a small
parameter. Let the phase function $S$ and $u$ in the
eikonal equation~(\ref{eikonal}) have the following behaviour as
$\epsilon \rightarrow 0$
\begin{eqnarray}
\label{y_direction}
&&S\left(x, \frac{\eta}{\epsilon},\xi\right) = \frac{\eta}{2 \epsilon^{2}} - \hat{S}\left(x,\eta,\xi \right),\nonumber \\
&&n^{2}\left(x, \frac{\eta}{\epsilon},\xi\right) = \frac{1}{4 \epsilon^{2}} -
q\left(x,\eta,\xi \right).
\end{eqnarray}
Considering the simplest cubic case~(\ref{cubic_case}) for $\alpha
= 1$ in Cartesian coordinates 
\begin{equation}
\label{csi_cartesian}
S_{\xi} = \frac{1}{4}S_{x}^{3} - \frac{3}{4} S_{x}S_{y}^{2} + U_{1}
S_{x} + U_{2} S_{y} 
\end{equation}
where $\gamma = U_{1} + i U_{2}$, and assuming that
\begin{equation}
U_{1} = \frac{3}{16 \epsilon^{2}} + \frac{3}{4} q,~~~~~~U_{2} = -2
\epsilon U, 
\end{equation}
one gets from~(\ref{eikonal}) and~(\ref{csi_cartesian}) the following equations
\begin{eqnarray}
\label{eikonal_KP}
\hat{S}_{\eta} &=& \hat{S}_{x}^{2} + q, \\
\label{eikonal_KP2}
\hat{S}_{\xi} &=& \hat{S}_{x}^{3} + \frac{3}{2} q \hat{S}_{x} + U.
\end{eqnarray}
Compatibility of equations~(\ref{eikonal_KP}) and~(\ref{eikonal_KP2}) gives rise to the well
known dispersionless Kadomtsev-Petviashvili (dKP) equation $q_{\xi} =
\frac{3}{2}q q_{x} + \frac{3}{4} \partial_{x}^{-1} q_{\eta \eta}$ and
$U = \frac{3}{4} \partial_{x}^{-1} q_{y}$. The dKP equation is rather well studied (see e.g.  \cite{Krichever}-\cite{Singular} and
reference therein). The KP equation itself represents the most known
2+1-dimensional integrable generalization of the KdV equation.

In the general case, assuming that the polynomial
$\varphi\left(x,\frac{\eta}{\epsilon},\xi;S_{x}, \epsilon S_{\eta}
\right)$ has an appropriate behavior as $\epsilon \rightarrow 0$, one
readily shows that equation~(\ref{csideformation}) is reduced to
\begin{equation}
\label{KP_hierarchy}
\hat{S}_{\xi} = \hat{\varphi}\left(x,\eta,\xi;\hat{S}_{x} \right)
\end{equation} 
where $\hat{\varphi}$ is an odd order polynomial in $\hat{S}_{x}$.
The condition of compatibility between equation~(\ref{eikonal_KP}) and
equations of the type~(\ref{KP_hierarchy}) gives rise to entire dKP
hierarchy. These 
equations describe propagation of the quasi-plane
wavefronts $y=const$ in a medium with ``very large'' refractive index.

The dVN and the dKP equations being relevant in particular situations
of propagations of waves in certain nonlinear media have an advantage
to be integrable.

\section{Quasiclassical $\bar{\partial}$-dressing method for the
  plane eikonal equation.}

In this and next sections we will demonstrate that the plane eikonal
equation and dVN hierarchy both for complex and real refractive indices
are treatable by the quasiclassical $\bar{\partial}$-dressing method.

The quasiclassical $\bar{\partial}$-dressing method is based on the
nonlinear Beltrami equation \cite{Konopelchenko2}-\cite{Konopelchenko4}
\begin{equation}
\label{Beltrami_nonlinear}
S_{\bar{\lambda}} = W\left(\lambda,\bar{\lambda};S_{\lambda} \right)
\end{equation}
where $S(\lambda,\bar{\lambda})$ is a complex valued function, $\lambda$ is the complex variable, $S_{\lambda} =
\der{S}{\lambda}$ and $W$ (the quasiclassical $\bar{\partial}$-data)
 is an analytic function of $S_{\lambda}$
\begin{equation}
\label{kernel}
W\left(\lambda,\bar{\lambda},S_{\lambda}\right) =
\sum_{n=0}^{\infty}w_{n}(\lambda,\bar{\lambda}) \left(S_{\lambda}\right)^{n},
\end{equation}
with some, in general, arbitrary functions
$w_{n}(\lambda,\bar{\lambda})$. 

To construct integrable equations one has to specify the domain $G$ (in
the complex plane $\mathbb{C}$) of
support for the function $W$ ($W=0$, $\lambda \in \mathbb{C}/G$) and look for solution of~(\ref{Beltrami_nonlinear})
in the form $S = S_{0} + \tilde{S}$, where the function $S_{0}$ is
analytic inside $G$, while $\tilde{S}$ is analytic outside $G$ \cite{Konopelchenko2}-\cite{Konopelchenko4}.
In order to construct the eikonal equation on the plane, we choose $G$ as
the ring ${\cal D} = \left \{\lambda \in \mathbb{C} :~ \frac{1}{a} <
|\lambda| < a \right\}$, where $a$ is an arbitrary real number ($a >
1$), and select $B-type$ solutions satisfying the constraint
\begin{equation}
\label{B_symmetry}
S\left(- \lambda,-\bar{\lambda}\right)= - S\left(\lambda,\bar{\lambda}\right).
\end{equation}
Then, we choose
\begin{equation}
\label{S_0}
S_{0} = z \lambda + \frac{\bar{z}}{\lambda}.
\end{equation}
Due to the analyticity of $\tilde{S}$ outside the ring and the
property~(\ref{B_symmetry}) one has
\begin{equation}
\tilde{S} = \sum_{n=0}^{\infty}
\frac{S_{2n+1}^{(\infty)}}{\lambda^{2n+1}};~~~~~\lambda \rightarrow \infty,
\end{equation}
and
\begin{equation}
\tilde{S} = \sum_{n=0}^{\infty}S_{2n+1}^{(0)}
\lambda^{2n+1}~~~~~\lambda \rightarrow 0.
\end{equation}
In particular, $\tilde{S}\left(0,0 \right) = 0$. \\
An important property of the nonlinear
$\bar{\partial}$-problem~(\ref{Beltrami_nonlinear}) is that the
derivatives $f = S_{t}$ of $S$ with respect to any independent
variable $t$, 
obeys the linear Beltrami equation
\begin{equation}
\label{Beltrami}
f_{\bar{\lambda}}
=W'\left(\lambda,\bar{\lambda};S_{\lambda}\right) f_{\lambda} 
\end{equation} 
where $W'\left (\lambda,\bar{\lambda};\phi \right) =
\der{W}{\phi}\left(\lambda,\bar{\lambda};\phi
\right)$. Equations~(\ref{Beltrami}) has two basic properties,
namely, 1) any differentiable function of solutions
$f_{1},\dots,f_{n}$ is again a 
solution; 2) under certain mild conditions on $W'$, a bounded solution
$f$ which is
equal to zero at certain point $\lambda_{0} \in \mathbb{C}$, vanishes
identically (Vekua's theorem) \cite{Vekua}.

These two properties allows us to construct an equation of the form
$\Omega\left(S_{z},S_{\bar{z}}\right) = 0$, with certain function $\Omega$. Indeed, taking into
account~(\ref{S_0}), one has
\begin{eqnarray}
S_{z}  = \lambda + \tilde{S}_{z}, \\
S_{\bar{z}} = \frac{1}{\lambda} + \tilde{S}_{\bar{z}},
\end{eqnarray}
i.e. $S_{z}$ has a pole at $\lambda = \infty $, while
$S_{\bar{z}}$ has a pole at $\lambda = 0$. The product $S_{z}
S_{\bar{z}}$ is again a solution of the linear Beltrami
equation~(\ref{Beltrami}) and it is bounded on the complex plane since
\begin{equation}
\label{bounded}
S_{z}S_{\bar{z}} = 1+ \frac{1}{\lambda} \tilde{S}_{z} + \lambda
\tilde{S}_{\bar{z}} + \tilde{S}_{z}\tilde{S}_{\bar{z}} 
\end{equation}
where $\tilde{S}_{\bar{z}}\left(\lambda=0 \right) = 0$ and $\lambda
\tilde{S}_{\bar{z}} \rightarrow S_{1\bar{z}}^{(\infty)}$ as $\lambda
\rightarrow \infty$.

Subtracting $1+S_{1\bar{z}}^{(\infty)}$ from the r.h.s of
equation~(\ref{bounded}), one gets a solution of
equation~(\ref{Beltrami}) which is bounded in $\mathbb{C}$ and
vanishes as $\lambda \rightarrow \infty$. According to the Vekua's
theorem it is equal to zero for all $\lambda$.  
Thus we get the equation
\begin{equation}
\label{eikonal2}
S_{z} S_{\bar{z}} = u(z,\bar{z})
\end{equation}
where
\begin{equation}
\label{residual}
u = 1 + S_{1,\bar{z}}^{(\infty)}.
\end{equation}
In the Cartesian coordinates $x,y$ defined by $z=x+iy$,
equation~(\ref{eikonal2}) is the standard two-dimensional eikonal
equation
\begin{equation}
\label{eikonal3}
\left(\nabla S \right)^{2} = n^{2}
\end{equation}
where $\nabla = \left(\der{}{x}, \der{}{y} \right)$ and $n^{2} =
\frac{u}{4}$.

Using the $\bar{\partial}$-problem~(\ref{Beltrami_nonlinear}), one
can, in principle,  construct solutions of
equation~(\ref{eikonal2}). So, the quasiclassical
$\bar{\partial}$-dressing method allows us to treat the plane eikonal
equation~(\ref{eikonal2}) in a way similar to dKP and d2DTL equations~\cite{Konopelchenko2}-\cite{Konopelchenko4}.
We note that the phase function $S$ in~(\ref{eikonal2}) depends also
on the complex variables $\lambda$ and $\bar{\lambda}$.
Curves $S(z,\bar{z})=const$ define
wavefronts. The $\bar{\partial}$-dressing approach provides us also with
the equation of light rays. Indeed, since r.h.s. of~(\ref{eikonal3}) does not
depend on $\lambda$ and $\bar{\lambda}$, the differentiation
of~(\ref{eikonal3}) with respect of $\lambda$ (or $\bar{\lambda}$)
gives
\begin{equation}
\label{scalar}
\nabla S \cdot \nabla \phi = 0
\end{equation}
where $\phi = S_{\lambda}$ (or $\phi = S_{\bar{\lambda}}$). So, the curves $S=const$ and
$\phi=const$ are reciprocally orthogonal and, hence, the latter ones
are nothing but the trajectories of propagating light. Thus,
the $\bar{\partial}$-dressing approach provides us with all
characteristics of the propagating light on the plane. Note that any
differentiable function $\phi \left(S_{\lambda}, S_{\bar{\lambda}}\right)$ is the
solution of equation~(\ref{scalar}) too.

Typically the refractive index is real and positive one being
defined in terms of high frequency limit of dielectric function and
magnetic permeability \cite{Born}. However, it was noted
in \cite{Veselago} that for certain media the product $\varepsilon_{0}
\mu_{0}$ and, hence, $u=4 n^{2}$ can be negative as well. Moreover, one
of the ways to describe the damping effects is to consider a
complex-valued $n^{2}$ \cite{Jackson,Veselago}. So, models of optical
phenomena deal both with real-valued and complex-valued refractive index.

In general, within the $\bar{\partial}$-dressing approach one has
a complex-valued phase function $S$ and, consequently, a complex
refractive index. 
To guarantee the reality of $u$, it is sufficient to impose
the following constraint on $S$
\begin{equation}
\label{parity_circle}
\overline{S}\left(\lambda,\bar{\lambda}\right) = S\left(\frac{1}{\bar{\lambda}},\frac{1}{\lambda} \right).
\end{equation}
Indeed, taking the complex conjugation of equation~(\ref{eikonal2}), using the
differential consequences (with respect to $z$ and $\bar{z}$), of the
above constraint and taking into account the independence of the
l.h.s. of equation~(\ref{eikonal2}) on $\lambda$,$\bar{\lambda}$, one
gets
\begin{equation}
\bar{u}\left (x_{n}\right) =
\bar{S}_{\bar{z}}\left(\lambda,\bar{\lambda} \right)
\bar{S}_{z}\left(\lambda,\bar{\lambda} \right) =
S_{\bar{z}}\left(\frac{1}{\bar{\lambda}},\frac{1}{\lambda} \right)
S_{z}\left(\frac{1}{\bar{\lambda}},\frac{1}{\lambda} \right) = u\left(x_{n} \right), 
\end{equation}
i.e. the ``refractive index'' $u$ is real one. The
constraint~(\ref{parity_circle}) leads to the relations
$\bar{S}_{2n+1}^{(0)} = S_{2n+1}^{(\infty)}$. 
Moreover, this constraint implies also that the function $S$ is real-valued
on the unit circle $\left|\lambda \right| = 1$ ($\overline{S}
\left(\lambda,\bar{\lambda} \right) =
S\left(\lambda,\bar{\lambda}\right)$, $|\lambda| =1$).
This provides us with the physical wavefronts. 

The $\bar{\partial}$-approach reveals also the connection between
geometrical optics and the theory of the, so-called, quasiconformal
mappings on the plane. Quasi-conformal mappings represents themselves a
very natural and important extension of the well-known conformal
mappings (see e.g. \cite{Ahlfors,Letho}). In contrast to the conformal mappings the
quasi-conformal  mappings are given by non-analytic functions, in
particular, by solutions of the Beltrami equation. 

According to \cite{Ahlfors,Letho} a solution of the nonlinear Beltrami
equation~(\ref{Beltrami_nonlinear}) defines a quasi-conformal mapping
of the complex plane $\mathbb{C}$. In our case we have a mapping
$S\left(\lambda,\bar{\lambda}\right)$ which is conformal outside the
ring $G$ and quasi-conformal inside $G$. Such mapping referred as the
conformal mapping with quasi-conformal extension. So, the
quasi-conformal mappings of this type which obey, in addition, the
properties~(\ref{B_symmetry}),~(\ref{S_0}) and~(\ref{parity_circle})
provide us with the solutions of the plane eikonal
equation~(\ref{eikonal2}). In particular, wavefronts given by
$S\left(\lambda,\bar{\lambda};z,\bar{z}\right) = const$ are level sets
of such quasi-conformal mappings. In more details, the interconnection
between quasi-conformal mappings and geometrical optics on the plane
will be discussed elsewhere.

\section{Quasiclassical $\bar{\partial}$-dressing method for dVN
  hierarchy.}
In this section we will apply the quasiclassical
$\bar{\partial}$-dressing method to the dVN hierarchy.
For this purpose we consider again the nonlinear
$\bar{\partial}$-problem~(\ref{Beltrami_nonlinear}), (\ref{kernel}) on
the ring $G$ with the constraints~(\ref{B_symmetry})
and~(\ref{parity_circle}) and introduce independent variables $x_{n}$,
$\bar{x}_{n}$  via
\begin{equation}
\label{asymp_S0}
S_{0} = \sum_{n=1}^{\infty}x_{n} \lambda^{2n-1} +
\sum_{n=1}^{\infty}\bar{x}_{n} \lambda^{-2n+1}.
\end{equation}
The derivatives $S_{x_{n}}$ and $S_{\bar{x}_{n}}$ obey the
linear Beltrami equation~(\ref{Beltrami}), and using the Vekua's
theorem one can construct an infinite set of equations of the form
\begin{equation}
\label{hierarchy}
\Omega\left(x_{n},\bar{x}_{n}, S_{x_{n}},S_{\bar{x}_{n}}
\right) = 0.
\end{equation} 
Repeating the procedure described in the previous section, one gets
the plane eikonal equation for $z=x_{1}$. In a similar manner, for the
variable $\xi = x_{2} = \bar{x}_{2}$, taking into account that
$S_{\xi} = \lambda^{3} + \frac{1}{\lambda^{3}} + \tilde{S}_{\xi}$, one
obtains the equation
\begin{equation}
\label{cubic}
S_{\xi} = S_{z}^{3}+S_{\bar{z}}^{3}+V S_{z} + \bar{V} S_{\bar{z}}
\end{equation}
where $V = - 3 S_{1z}^{(\infty)} = - 3 \partial_{\bar{z}}^{-1}u_{z}$.
Evaluating the terms of the order $\lambda^{-1}$ in the both sides of
equation~(\ref{cubic}), one gets the dVN equation
\begin{equation}
\label{dVN}
u_{\xi} = - 3 \left(u \partial_{\bar{z}}^{-1} u_{z} \right)_{z} - 3
\left(u \partial_{z}^{-1}u_{\bar{z}} \right)_{\bar{z}}.
\end{equation}
Considering the higher variables $\xi_{n} = x_{n} = \bar{x}_{n}$, ($n
= 2, 3, \dots$), one constructs all
equations~(\ref{higherpolynomials}), and hence, the whole dVN
hierarchy.
It is a straightforward check that the
constraint~(\ref{parity_circle}) is compatible with equations~(\ref{cubic})
and~(\ref{higherpolynomials}). 
So, the $\bar{\partial}$-dressing scheme under this constraint
provides us with the real-valued solutions of the dVN hierarchy.

If one relaxes the reality condition for $u$ (i.e. does not impose the
constraint~(\ref{parity_circle})), then there is a wider family of
integrable deformations of the plane eikonal equation with complex
refractive index. To build these deformations one again considers the
$\bar{\partial}$-problem~(\ref{Beltrami_nonlinear}), chooses the domain
$G$ as the ring (as before), imposes the
$B$-constraint~(\ref{B_symmetry}), but now chooses $S_{0}$ as follows
\begin{equation}
\label{asymp_S0_general}
S_{0} = \sum_{n=1}^{\infty}x_{n} \lambda^{2n-1} +
\sum_{n=1}^{\infty} y_{n} \lambda^{-2n+1}
\end{equation}
where $y_{1} = \bar{x}_{1} = \bar{z}$. Repeating the previous construction,
one gets again the eikonal equation~(\ref{eikonal2}), but now with a
complex-valued $u$. Considering the derivatives $S_{x_{n}}$,
$S_{y_{n}}$ with $n=2,3,\dots$, one obtains the two set of
equations
\begin{eqnarray}
\label{higher_x}
&&S_{x_{n}}=\sum_{m=1}^{n} u_{m}(x_{n},y_{n}) \left(S_{z}
\right)^{2m-1},~~~~~~n = 2,3,\dots \\
\label{higher_y}
&&S_{y_{n}}=\sum_{m=1}^{n} v_{m}(x_{n},y_{n}) \left(S_{\bar{z}}
\right)^{2m-1},~~~~~~n = 2,3,\dots 
\end{eqnarray}
Equations~(\ref{eikonal2}) and~(\ref{higher_x}) give rise to the
hierarchy of equations
\begin{equation}
\label{higher_x_2}
u_{x_{n}} = F_{n}\left(u,u_{z},u_{\bar{z}}\right),
\end{equation}
the simplest of which is of the form
\begin{equation}
\label{dVN_x}
u_{x_{2}} = -3\left(u \partial_{\bar{z}}^{-1}u_{z}\right)_{z}
\end{equation} 
where
\begin{equation}
S_{x_{2}} = S_{z}^{3} + V S_{z}.
\end{equation}
Equations~(\ref{eikonal2}) and~(\ref{higher_y}) generate the hierarchy
of equations $u_{y_{n}} = \tilde{F}_{n}$, the simplest of which
is given by
\begin{equation}
\label{dVN_y}
u_{y_{2}} = - 3 \left(u
\partial_{z}^{-1}u_{\bar{z}}\right)_{\bar{z}},
\end{equation} 
and
\begin{equation}
S_{y_{2}} = S_{\bar{z}}^{3} + \bar{V} S_{\bar{z}}.
\end{equation}
For both of these hierarchies the quantity $\int \int_{\mathbb{C}} u dz
\wedge d\bar{z}$ is the integral of motion as for the dVN hierarchy. It
is easy to see that equations~(\ref{dVN_x}) and~(\ref{dVN_y}) imply
the dVN equation~(\ref{dVN}) for the variable $\xi = \frac{x_{2}+y_{2}}{2}$.

\section{Characterization of $\bar{\partial}$-data.}
As we have seen, the constraints~(\ref{B_symmetry})
and~(\ref{parity_circle}) guarantee that one will get the eikonal
equation~(\ref{eikonal2}) with real valued $u$. In this section we
will discuss the characterization conditions for $\bar{\partial}$-data
$W\left(\lambda,\bar{\lambda}, S_{\lambda}\right)$ which provide us
with such result.
In general, if one considers $\bar{\partial}$-problem with a kernel
defined in a ring and the function $S_{0}$ singular at two points,
e.g. $\lambda =0$ and $\lambda = \infty$, without
$B$-constraints~(\ref{B_symmetry}), one constructs the dispersionless
Laplace hierarchy \cite{Konopelchenko4} associated with the quasiclassical limit of the
Laplace equation, i.e. with the equation

\begin{equation}
\label{magnetic}
S_{z}(\lambda,z,\bar{z}) S_{\bar{z}}(\lambda,z,\bar{z}) - a S_{\bar{z}} =
u(z,\bar{z}),~~~~ \forall \lambda \in \mathbb{C} 
\end{equation} 
where $a = \partial_{z}\tilde{S}(0,0)$ and $u = 1 + \partial_{\bar{z}}S_{1}^{(\infty)}$.
The eikonal equation~(\ref{eikonal}) is obtained as a reduction of
equation~(\ref{magnetic}) taking $\tilde{S}(0,0)$ independent on
$z$. In fact, the $B$ condition~(\ref{B_symmetry}), producing
$\tilde{S}(0,0)=0$ realizes this reduction. In what follows we will
discuss how one has to choose the $\bar{\partial}$-data in a way to
construct the dVN hierarchy directly. 
We will find  constraints which are dispersionless analog of the
constraints found in \cite{Grinevich}, which specify two-dimensional
Schr\"odinger equation with real-valued potential.
In particular, we will see that it is possible to weaken slightly the
$B$-condition~(\ref{B_symmetry}), 
since the value $\tilde{S}(0,0)$ is fixed up to a constant by dVN reduction.  

In the following we will focus on solutions of $\bar{\partial}$
problem of the form $S = S_{0}+\tilde{S}$, where $S_{0}$ has
polynomial singularities at $\lambda = 0$ and $\lambda = \infty$ and
$\tilde{S}$ is holomorphic at these points and such that
\begin{eqnarray}
\label{infinity_limit}
&& \lim_{\lambda = \infty}\tilde{S}\left(\lambda,\bar{\lambda}\right)
  = 0, \\ 
\label{zero_limit}
&& \lim_{\lambda = 0}\tilde{S}\left(\lambda,\bar{\lambda}\right)
  = \tilde{S}(0,0,z,\bar{z}) = -i v(z,\bar{z}), 
\end{eqnarray}

{\lemma 
\label{lemma1}
Let the kernel $W$ in equation~(\ref{Beltrami_nonlinear}) satisfies the assumptions
of the Vekua's theorem, and let $S$ be its solution. The condition
\begin{equation}
\label{parity_S}
S\left(\lambda,\bar{\lambda}\right) =
-S\left(-\lambda,-\bar{\lambda}\right) + const
\end{equation}
is verified if and only if
\begin{equation}
\label{parity_W}
W'\left(\lambda,\bar{\lambda},S_{\lambda}\left(\lambda,\bar{\lambda}
\right) \right) = W'\left(-\lambda,-\bar{\lambda},-S_{\lambda}\left(-\lambda,-\bar{\lambda}
\right) \right)
\end{equation}}

{\proof
The condition~(\ref{parity_W}) is a necessary one. Indeed, starting
from~(\ref{Beltrami_nonlinear}), one obtains

\begin{eqnarray}
\label{first_one}
\der{}{\bar{\lambda}}\left(S_{z}
\left(\lambda,\bar{\lambda} \right) \right) =
W'\left(\lambda,\bar{\lambda}, S_{\lambda}\left(\lambda,\bar{\lambda}\right)\right)
\der{}{\lambda}\left(S_{z}\left(\lambda,\bar{\lambda}\right)
\right) \\
\label{last_one}
\der{}{\bar{\lambda}}\left(S_{z}
\left(-\lambda,-\bar{\lambda} \right) \right) =
W'\left(-\lambda,-\bar{\lambda}, -S_{\lambda}\left(-\lambda,-\bar{\lambda} \right)\right)
\der{}{\lambda}\left(S_{z}\left(-\lambda,-\bar{\lambda}\right)
\right) 
\end{eqnarray}
Exploiting the condition~(\ref{parity_S}) in~(\ref{last_one}), one gets
the equality~(\ref{parity_W}).

The condition~(\ref{parity_W}) is sufficient. Indeed, let us introduce
the function 
\begin{equation}
\label{Phi}
\Phi\left(\lambda,\bar{\lambda},z,\bar{z}\right)  =
S_{z}\left(\lambda,\bar{\lambda}\right) +
S_{z}\left(-\lambda,-\bar{\lambda} \right).
\end{equation}
Using the equations~(\ref{infinity_limit})
and~(\ref{zero_limit}), one has
\begin{eqnarray}
&&\lim_{\lambda \rightarrow
    \infty}\Phi\left(\lambda,\bar{\lambda},z,\bar{z} \right) =
    0, \\
&&\lim_{\lambda \rightarrow
    0}\Phi\left(\lambda,\bar{\lambda},z,\bar{z} \right) = -2 i
    v_{z}\left(z,\bar{z} \right).
\end{eqnarray}
Both terms in the right hand side of~(\ref{Phi}) satisfy the Beltrami
equation~(\ref{first_one}) and~(\ref{last_one}) respectively, 
from which, exploiting the constraint~(\ref{parity_W}), one concludes that
\begin{equation}
\der{\Phi}{\bar{\lambda}}\left(\lambda,\bar{\lambda},z,\bar{z}
\right) =
W'\left(\lambda,\bar{\lambda},S_{\lambda}\left(\lambda,\bar{\lambda}
\right) \right) \der{\Phi}{\lambda}\left(\lambda,\bar{\lambda},z,\bar{z} \right).
\end{equation}
The function $\Phi$ is a solution of the Beltrami equation vanishing
at $\lambda \rightarrow \infty$. So, $\Phi$ vanishes identically on
whole $\lambda$-plane, so
\begin{equation}
\label{identity1}
\der{S}{z}\left(\lambda,\bar{\lambda}\right) = -
\der{S}{z}\left(-\lambda,-\bar{\lambda} \right).
\end{equation}
In particular
$\Phi\left(0,0,z,\bar{z} \right)= -2 i v_{z}\left(
z,\bar{z}\right) \equiv 0$ ,that is $v_{z}\left(z,\bar{z}\right)
\equiv 0$. 
Analogously it is possible to demonstrate
that
\begin{equation}
\label{identity2}
\der{S}{\bar{z}}\left(\lambda,\bar{\lambda}\right) = -
\der{S}{\bar{z}}\left(-\lambda,-\bar{\lambda} \right).
\end{equation}
Hence $v_{\bar{z}}\left(
z,\bar{z}\right) \equiv 0$. Equations~(\ref{identity1})
and~(\ref{identity2}) lead to the relation~(\ref{parity_S}) $\Box$

{\lemma  
\label{lemma2}
Let $S\left(\lambda,\bar{\lambda},z,\bar{z} \right)$ be a
solution of $\bar{\partial}$ problem~(\ref{Beltrami_nonlinear}) such
that~(\ref{infinity_limit}) and~(\ref{zero_limit})  are verified. Then the condition
\begin{equation}
\label{unit_inversion}
\bar{S}\left(\lambda,\bar{\lambda}\right) =
  S\left(\frac{1}{\bar{\lambda}}, \frac{1}{\lambda} \right)+iv(z,\bar{z})
\end{equation}
is verified if and only if 
\begin{equation}
\label{kernel_inversion}
 \lambda^{2}
  \overline{W}\left(\lambda,\bar{\lambda},
  S_{\lambda}\left(\lambda,\bar{\lambda} \right) \right) = -
  W\left(\frac{1}{\bar{\lambda}},\frac{1}{\lambda}, -
  \bar{\lambda}^{2}
  S_{\bar{\lambda}}\left(\frac{1}{\bar{\lambda}},\frac{1}{\lambda}
  \right) \right).
\end{equation}}

{\proof The condition~(\ref{kernel_inversion}) is a necessary one. Let
  us consider the complex conjugation of equation~(\ref{Beltrami_nonlinear})
\begin{equation}
\label{DBAR_conjugate}
\partial_{\lambda}\left(\bar{S}\left(\lambda,\bar{\lambda} \right)\right) =
\overline{W}\left(\lambda,\bar{\lambda}, S_{\lambda}\left(\lambda,\bar{\lambda} \right) \right).
\end{equation} 
Since 
\begin{equation}
\bar{S}\left(\lambda,\bar{\lambda} \right) =
S\left(\frac{1}{\bar{\lambda}},\frac{1}{\lambda} \right) + i
v(z,\bar{z}) = S\left(\xi,\bar{\xi} \right) + i v(z,\bar{z})
\end{equation}
where $\xi = \bar{\lambda}^{-1}$ and $\bar{\xi} = \lambda^{-1}$, the
left hand side of~(\ref{DBAR_conjugate}) can be written as follows
\begin{eqnarray}
\partial_{\lambda}\left(S\left(\xi,\bar{\xi} \right) \right) &=&
\der{\bar{\xi}}{\lambda} \der{S}{\bar{\xi}}\left(\xi,\bar{\xi} \right)=-\frac{1}{\lambda^{2}}
W\left(\xi,\bar{\xi},\partial_{\xi}S\left(\xi,\bar{\xi} \right)
\right) = \nonumber \\
&=& - \frac{1}{\lambda^{2}} W\left(\frac{1}{\bar{\lambda}},
\frac{1}{\lambda}, -\bar{\lambda}^{2} \partial_{\bar{\lambda}}S\left(\frac{1}{\bar{\lambda}},\frac{1}{\lambda} \right) \right),
\end{eqnarray}
that provides us with equation~(\ref{unit_inversion}). 

The condition~(\ref{kernel_inversion}) is a sufficient one.
The $\bar{\partial}$-equation~(\ref{Beltrami_nonlinear}) written in
terms of the variables $\xi = \bar{\lambda}^{-1}$ and $\bar{\xi} =
\lambda^{-1}$  
\begin{equation}
S_{\bar{\xi}} = W\left(\xi,\bar{\xi},S_{\xi}\left(\xi,\bar{\xi} \right) \right)
\end{equation}
is equivalent to
\begin{equation}
\label{DBAR_equiv}
\lambda^{2} S_{\lambda}\left(\frac{1}{\bar{\lambda}},\frac{1}{\lambda}
\right) = -
W\left(\frac{1}{\bar{\lambda}},\frac{1}{\lambda},-\bar{\lambda}^{2}S_{\bar{\lambda}}\left(\frac{1}{\bar{\lambda}},\frac{1}{\lambda} \right) 
\right).
\end{equation}
Using equation~(\ref{DBAR_conjugate}), multiplied by $\lambda^{2}$, and
equation~(\ref{kernel_inversion}), one concludes that
\begin{equation}
\label{parity_circle_der}
\partial_{\lambda}\bar{S}\left(\lambda,\bar{\lambda}\right) =
\partial_{\lambda} S\left(\frac{1}{\bar{\lambda}},\frac{1}{\lambda} \right).
\end{equation} 
Integrating~(\ref{parity_circle_der}), one gets
\begin{equation}
\label{jump}
\bar{S}\left(\lambda,\bar{\lambda} \right) =
S\left(\frac{1}{\bar{\lambda}},\frac{1}{\lambda} \right)+
\tilde{v}\left(z,\bar{z} \right).
\end{equation}

Let us note that $\tilde{v}$ cannot depend on $\bar{\lambda}$ since
the function $S$ is meromorphic outside the ring.
Now, evaluating the equality~(\ref{jump}) at $\lambda=0$ and $\lambda
\rightarrow \infty$, one obtains
\begin{equation}
\tilde{v}\left(z,\bar{z} \right) = \overline{\tilde{S}}(0,0,z,\bar{z}) =
- \tilde{S}(0,0,z,\bar{z}).
\end{equation}
So, $\tilde{v}$ is a purely imaginary function
\begin{equation}
\tilde{v}\left(z,\bar{z}\right) = i v\left(z,\bar{z} \right)
\end{equation}
where $v$ is a real valued function. This complete the proof $\Box$.

Note that $\tilde{v}_{z} = i v_{z}(z,\bar{z}) = -a$,
in other words, it is the coefficient in front of the ``magnetic'' term in
equation~(\ref{magnetic}). When it vanishes (i.e. $a=0$), one has the
pure potential equation~(\ref{magnetic}), that is the
eikonal equation.

Combining together the lemmas~(\ref{lemma1}) and~(\ref{lemma2}), one gets the
following theorem:

{\bf Theorem} {\em If the $\bar{\partial}$-data $W$ of the
  $\bar{\partial}$-equation~(\ref{Beltrami_nonlinear}) obey the constraints
\begin{eqnarray}
\lambda^{2}
\overline{W}\left(\lambda,\bar{\lambda},S_{\lambda}\left(\lambda,\bar{\lambda}
\right) \right) = -
W\left(\frac{1}{\bar{\lambda}},\frac{1}{\lambda},-\bar{\lambda}^{2}S_{\bar{\lambda}}\left(\frac{1}{\bar{\lambda}},\frac{1}{\lambda}\right)
\right), \\
W'\left(\lambda,\bar{\lambda},S_{\lambda}\left(\lambda,\bar{\lambda}\right)
\right) = W'\left(-\lambda,-\bar{\lambda},-S_{\lambda}\left(-\lambda,-\bar{\lambda}\right)\right),
\end{eqnarray}
then this $\bar{\partial}$-problem provides us with the eikonal equation
with real-valued refractive index.
}

\section{Symmetries, symmetry constraints and reduction method for dVN
 equation.}
The dVN equation and  the dVN hierarchy possess a number of
remarkable properties typical for the dispersionless integrable
equations and hierarchies \cite{Konopelchenko4,KonopMoro}. Here we will discuss some
aspects of symmetry properties of the complex dVN equation, symmetry
constraints and corresponding $1+1$-dimensional equations which
provides us with solutions of the eikonal equation and the dVN
equation.

We will concentrate on equation~(\ref{dVN_x}), i.e. equation
\begin{equation*}
u_{x_{2}} = -3 \left(u \partial_{\bar{z}}^{-1} u_{z} \right)_{z},
\end{equation*}
 with the
complex-valued $u$, which is the compatibility condition for the system
\begin{eqnarray}
\label{system_a}
&&S_{z} S_{\bar{z}} = u, \\
\label{system_b}
&&S_{x_{2}} = S_{z}^{3} + V S_{z}
\end{eqnarray}
where $V_{\bar{z}} = -3 u_{z}$.
Infinitesimal continuous symmetries $u \rightarrow u + \delta u$ of
equation~(\ref{dVN_x}) are defined, as usual, by its linearized
version, i.e. by the equation
\begin{equation}
\label{dVN_linearized}
{\cal L} \delta u = 0
\end{equation}
where the linear operator ${\cal L}$ acts as follows
\begin{equation*}
{\cal L} = \partial_{x_{2}} + 3
\partial_{z}\left(\left(\partial_{\bar{z}}^{-1}u_{z}\right) (\cdot)
\right) + 3 \partial_{z} \left(u \partial_{\bar{z}}^{-1} \partial_{z} (\cdot)    \right).
\end{equation*}
The system~(\ref{system_a}-\ref{system_b}) is quite relevant for an analysis of
equation~(\ref{dVN_linearized}). Namely, it is straightforward to check
that a class of its solutions is given by
\begin{equation}
\label{symmetry1}
\delta u  = \sum_{i=1}^{N} c_{i} \left(S_{i} - \tilde{S}_{i} \right)_{z\bar{z}}
\end{equation}  
where $S_{i}$, $\tilde{S}_{i}$ are arbitrary solutions of the
system~(\ref{system_a}), (\ref{system_b}) with given $u$, $c_{i}$ are arbitrary constants
and $N$ is an arbitrary integer.
 
The formula~(\ref{symmetry1}) provides us with a wide class of
symmetries of the dVN equation. In particular, one can choose $S_{i} =
S\left(\lambda=\lambda_{i} \right)$,
and $\tilde{S}_{i} =
\tilde{S}\left(\lambda=\tilde{\lambda}_{i}\right)$. In the case
$\tilde{\lambda}_{i}= \lambda_{i} + \mu_{i}$, where $\mu_{i}
\rightarrow 0$ and $c_{i} = \frac{\tilde{c}_{i}}{\mu_{i}}$, one has a
class of symmetries given by
\begin{equation}
\label{symmetry2}
\delta u = \sum_{i=1}^{N} \tilde{c}_{i} \phi_{iz\bar{z}}
\end{equation}
where $\phi_{i} = \left
. \der{S}{\lambda}(\lambda)\right|_{\lambda = \lambda_{i}}$. In the
simplest case $n=1$ one has 
\begin{equation}
\delta u = c \phi_{z\bar{z}},
\end{equation}
i.e. a symmetry is given by the divergence of the vector
$\left(\phi_{x},\phi_{y} \right)$, which is normal to the light rays
$\phi = const$, or tangent to the wavefronts $S = const$. 

Note that
one has the same class of symmetries also for the real dVN
equation~(\ref{dVN}).  

Any linear superposition of infinitesimal symmetries is obviously a
symmetry too. In virtue of~(\ref{dVN_linearized}), the requirement
that certain linear superposition of symmetries vanishes, is apparently
compatible with the dVN equation.
An obvious set of symmetries of the dVN equation is given by $u_{x_{n}}$, $u_{y_{n}}$. Combining them with the
symmetries~(\ref{symmetry1}), one gets the following set of possible
symmetry constraints
\begin{equation}
\label{set_symmetry}
\sum_{n} \left (\alpha_{n} u_{x_{n}} + \beta_{n} u_{y_{n}} \right) =
\sum_{i=1}^{N} c_{i} \left(S_{i} - \tilde{S}_{i} \right)_{z\bar{z}}
\end{equation}   
where $\alpha_{n}$ and $\beta_{n}$ are arbitrary constants.
Symmetry constraints~(\ref{set_symmetry}) are compatible with the dVN
hierarchy and reduce it to the families of the $1+1$-dimensional
equations similar to the dKP and 2DTL case \cite{Bogdanov1}.

Here we will consider the simplest symmetry constraint of the form
\begin{equation}
u_{z} = \phi_{z\bar{z}}.
\end{equation} 
Assuming that the integration ``constant'' is equal to zero, one has 
\begin{equation}
u = \phi_{\bar{z}}.
\end{equation}
Equations~(\ref{system_a}) and~(\ref{system_b}) imply
\begin{eqnarray}
\label{symmetry3_a}
&&S_{z} S_{\bar{z}} = \phi_{\bar{z}}, \\
\label{symmetry3_b}
&&\phi_{z}S_{\bar{z}} + S_{z} \phi_{\bar{z}} = 0, \\
\label{symmetry3_c}
&&\phi_{x_{2}} = 3 S_{z}^{2} \phi_{z} + V \phi_{z}
\end{eqnarray}
and
\begin{equation}
V = -3 \phi_{z}.
\end{equation}
Substitution of~(\ref{symmetry3_a}) into~(\ref{symmetry3_b}) gives
\begin{equation}
\label{symmetry4}
\phi_{z} = - S_{z}^{2}.
\end{equation}
The compatibility condition for equations~(\ref{symmetry3_a})
and~(\ref{symmetry4}), readily, leads to the equation
\begin{equation}
S_{z} S_{\bar{z}}^{3} = \alpha^{3}
\end{equation}
where $\alpha$ is an arbitrary constant. So 
\begin{equation}
\label{symmetry5}
S_{\bar{z}} = \alpha S_{z}^{-\frac{1}{3}},
\end{equation}
while equation~(\ref{symmetry3_a}) becomes
\begin{equation}
\label{symmetry6}
\phi_{\bar{z}} = \alpha i^{-\frac{2}{3}} \phi_{z}^{\frac{1}{3}}.
\end{equation}
Equations~(\ref{system_b}) and~(\ref{symmetry3_c}) take the
form
\begin{eqnarray}
\label{system2_a}
&&S_{x_{2}} = 4 S_{z}^{3} \\
\label{system2_b}
&&\phi_{x_{2}} = - 6 \phi_{z}^{2}.
\end{eqnarray}
It is easy to check that equations~(\ref{symmetry6})
and~(\ref{system2_b}) are satisfied in virtue of
equations~(\ref{symmetry5}),~(\ref{system2_a}) and the
relation~(\ref{symmetry4}). 
Thus, we have proved that any common solution $S(z,\bar{z},x_{2})$  of
equations 
\begin{eqnarray}
\label{system3_a}
&&S_{\bar{z}} = \alpha S_{z}^{-\frac{1}{3}} \\
\label{system3_b}
&&S_{x_{2}} = 4 S_{z}^{3},
\end{eqnarray}
provides us with a solution of the complex dVN
equation~(\ref{dVN_x}), given by $u =\alpha S_{z}^{\frac{2}{3}}$
and $V = 3 S_{z}^{2}$. Of course, any solution $S(z,\bar{z})$ of
equation~(\ref{system3_a}) gives us a solution of the eikonal
equation~(\ref{system_a}) with the refractive index $u = \alpha
S_{z}^{\frac{2}{3}}$.

One can rewrite equations~(\ref{system3_a}) and~(\ref{system3_b}) in
the form of quasilinear equations for $\psi = S_{z}$ or directly in
terms of $u = \alpha S_{z}^{\frac{2}{3}}$. One has 
\begin{eqnarray}
\label{quasilinear_a}
&&u_{\bar{z}} = - \frac{\alpha}{3} u^{-2} u_{z}, \\
\label{quasilinear_b}
&&u_{x_{2}} = 12 \alpha^{-3} u^{3}u_{z}.
\end{eqnarray}
It is an easy check that a function $u$ which obeys both these
equations is a solution of equation~(\ref{dVN_x}).
Equations~(\ref{quasilinear_a}) and~(\ref{quasilinear_b}) are
$1+1$-dimensional quasilinear equations. So, the problem of
construction of solutions of the $2+1$-dimensional dVN equation is
reduced to the 
construction of common solutions of $1+1$-dimensional equations. Such
a method of construction of solutions for dispersionless integrable
equations is known as the reduction method \cite{Krichever,Kodama,TakTak,Bogdanov1,Ferapontov,Manas,Chang,Baldwin}.

The method of characteristics provides us with general solution of
equations~(\ref{quasilinear_a}) and~(\ref{quasilinear_b}) in the
hodograph form
\begin{equation}
\label{hodograph}
z - \frac{\alpha^{3}}{3} u^{-2} \bar{z} + 12 \alpha^{-3} u^{3} x_{2} -
\psi(u) =0 
\end{equation}
where $\psi(u)$ is an arbitrary function. Solutions of
equation~(\ref{hodograph}) give us a class of solutions of the
complex dVN equation~(\ref{dVN_x}).
In a similar manner one can treat the symmetry constraint $u_{\bar{z}}
= \phi_{z\bar{z}}$, for equation~(\ref{dVN_y}). Namely, one gets
equations~(\ref{system3_a}-\ref{hodograph}) with the substitution $z
\leftrightarrow \bar{z}$ and $x_{2} \leftrightarrow y_{2}$.

Under more general constraint
\begin{equation}
\label{symmetry_gen}
u_{z} = \sum_{i=1}^{N} c_{i} \left(S_{i}-\tilde{S}_{i} \right)_{z\bar{z}},
\end{equation}
one has
\begin{eqnarray}
\label{symmetry_gen_a}
&&u = \sum_{i=1}^{N} c_{i} \left(S_{i}-\tilde{S}_{i}
  \right)_{\bar{z}}\\
\label{symmetry_gen_b}
&&V = -3 \sum_{i=1}^{N} c_{i} \left(S_{i}- \tilde{S}_{i} \right)_{z},
\end{eqnarray}
and equations~(\ref{system_a}) and~(\ref{higher_x}) at $n=2$ are reduced to the
following systems
\begin{eqnarray}
\label{system_gen_1_a}
&&S_{kz}  S_{k\bar{z}} = \sum_{i=1}^{N} c_{i} \left(S_{i} -
\tilde{S}_{i} \right)_{\bar{z}},  \\ 
\label{system_gen_1_b}
&&\tilde{S}_{kz}  \tilde{S}_{k\bar{z}} = \sum_{i=1}^{N} c_{i} \left(S_{i} -
\tilde{S}_{i} \right)_{\bar{z}},~~~~~~k = 1,\dots,N,  
\end{eqnarray}
and 
\begin{eqnarray}
\label{system_gen_2_a}
&&S_{k x_{2}} = S_{k z}^{3} -3 S_{k z} \sum_{i=1}^{N}c_{i}
\left(S_{i}-\tilde{S}_{i} \right)_{z} \\
\label{system_gen_2_b}
&&\tilde{S}_{k x_{2}} = \tilde{S}_{k z}^{3} -3 \tilde{S}_{k z} \sum_{i=1}^{N}c_{i}
\left(S_{i}-\tilde{S}_{i} \right)_{z}~~~~~~k = 1,\dots,N. 
\end{eqnarray}
Common solutions of equations~(\ref{system_gen_1_a}-\ref{system_gen_2_b}) provide us with solutions of the complex
dVN equation~(\ref{dVN_x}) given by~(\ref{symmetry_gen_a}).

These type of systems, symmetry constraints for the real dVN
equation~(\ref{dVN}) and associated constructions within the
quasiclassical $\bar{\partial}$-approach will be discussed in the
paper \cite{BKM}.

\section{Examples of solutions: wavefronts and refractive indices.}
In this section we shall present, as illustrative examples, several
exact and numerical solutions of the eikonal equation, the dVN and dKP
equations and visualize wavefronts distribution and refractive indices. For this end, we will use the solutions of dKP
hierarchy obtained by quasiclassical $\bar{\partial}$-dressing
approach in \cite{Solutions}, and solutions
of dVN hierarchy obtained in the previous section.

\subsection{dKP wavefronts.}
In order to interpret the $S$ function associated with the dKP equation  as a perturbation of a plane
wavefront, it is convenient to rewrite equations~(\ref{y_direction}) as follows
\begin{eqnarray}
\beta S = \eta - \beta \hat{S}, \\
\beta n^{2} = \frac{1}{2} - \beta q
\end{eqnarray}
where $\beta = 2 \epsilon^{2}$ is a small parameter, and $\hat{S}$ satisfies
equations~(\ref{eikonal_KP}) and~(\ref{eikonal_KP2}). In what follows
we will discuss a two-dimensional explicit solution and its
three-dimensional generalization.
In our interpretation, the two-dimensional solution describes
wavefronts  which are plane, that is, they do not depend on $\xi$
variable.
The three-dimensional solutions show us slow deformations
along $\xi$-axis. 

The purely two-dimensional case is given by
\begin{eqnarray}
\label{S_solution_2D}
&&\hat{S} = \frac{1}{2} \frac{\left(\lambda -b\right)^{2}}{a - 2
\bar{\lambda}} - c;~~~~~\left| \lambda \right| \leq 1,  \\
\label{dkp_index_2D}
&&q = - \left(4 x + 16 \eta^{2} \right)
\end{eqnarray}
where
\begin{equation}
a = \frac{1}{\eta},~~~~~b = 2 \eta - \frac{x}{2 \eta},~~~~~c =
\frac{\left(x + 4 \eta^{2}  \right)^{2}}{4 \eta}.
\end{equation}
It is associated with the solution obtained in the paper \cite{Solutions} for the particular choice of the kernel
$W = S_{\lambda}^{2}$.
Calculating $\hat{S}$ at $\lambda = 0$, we obtain a real and non-singular
phase function on the $x,\eta$-plane
\begin{equation}
\label{phase_2D}
\hat{S}(\lambda = 0) = - 4 x \eta.
\end{equation}
The equation of the corresponding wavefronts is the following
\begin{equation}
\label{dkp_wfront_2D}
x = \frac{k - \eta}{4 \beta \eta}
\end{equation}
where $k$ is an arbitrary constant, whose range of validity is restricted to the region where $\left|
\frac{\beta \hat{S}}{\eta} \right| \ll 1$. In this case it is the strip
$\left|x\right| \ll \frac{1}{4 \beta}$.
The figure~\ref{fig_dkpfront}-a) shows the distribution of wavefronts
(obtained setting different values of the parameter $k$), and the
shady region represent the strip $\left|x\right| \leq \frac{1}{8 \beta}$
where the approximation reasonably holds. The
figure~\ref{fig_dkpindex} -b) represents the refractive index density.
Light rays turn off towards darker regions corresponding to larger
refractive index. 
\begin{figure}
\includegraphics[width=4.5cm]{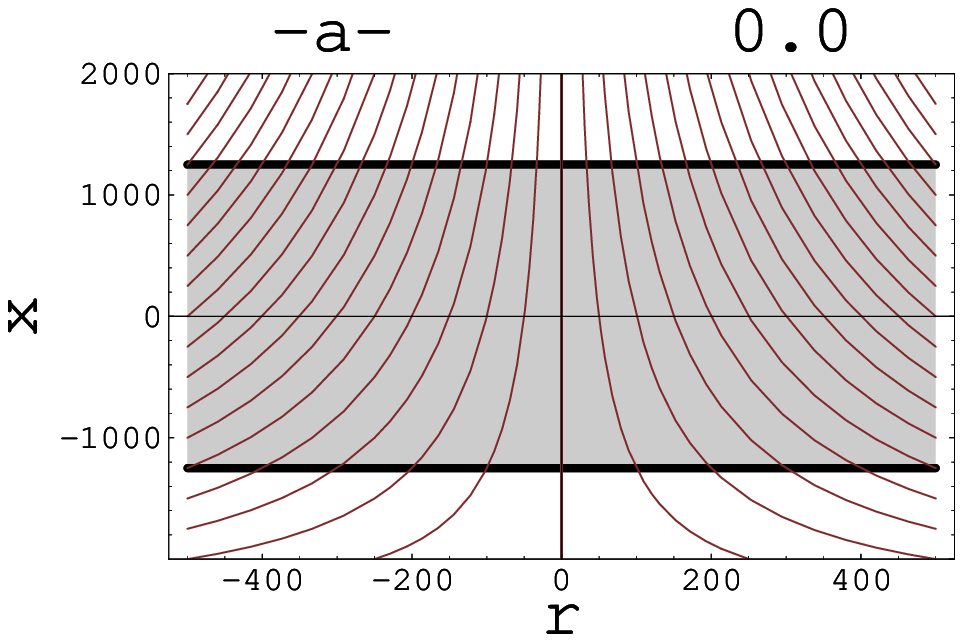}
\includegraphics[width=4.5cm]{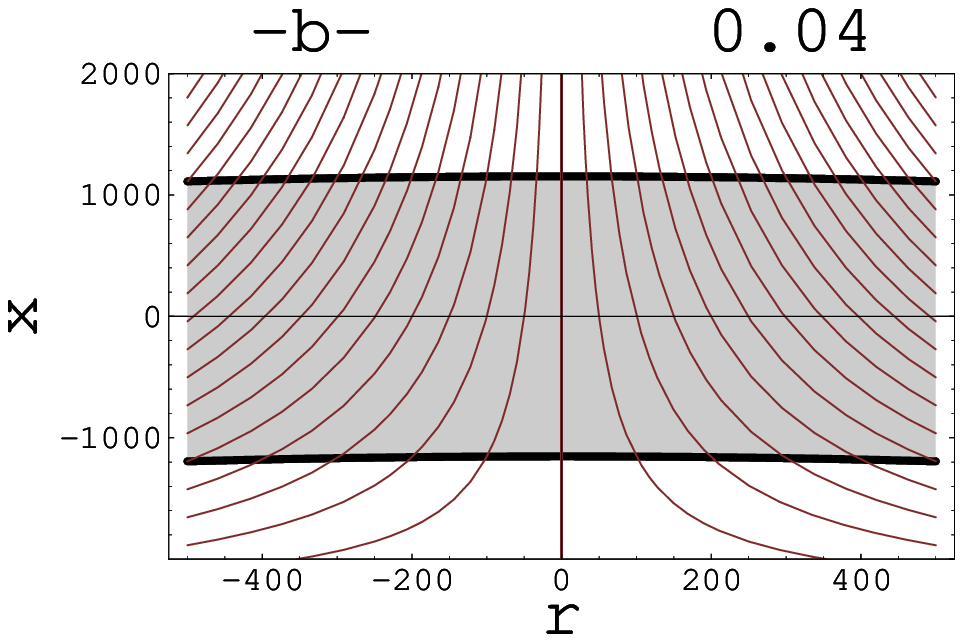}
\includegraphics[width=4.5cm]{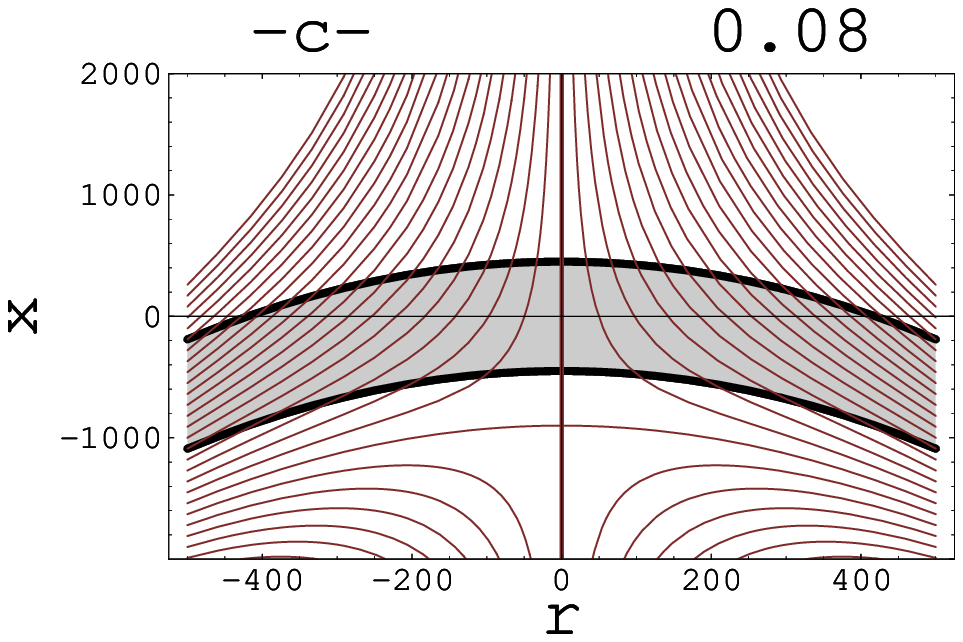}
\centerline{\includegraphics[width=4.5cm]{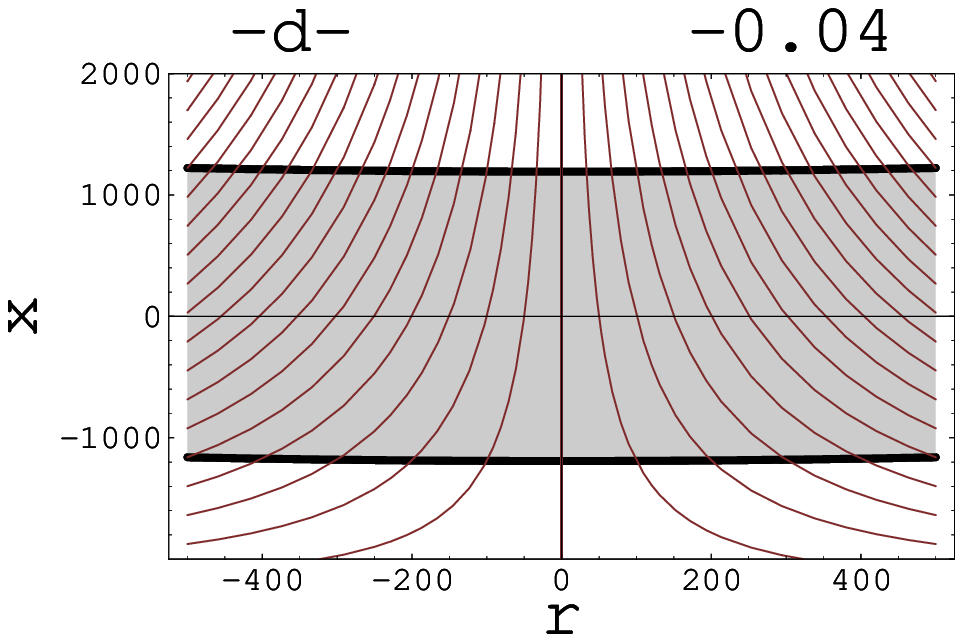}
\includegraphics[width=4.5cm]{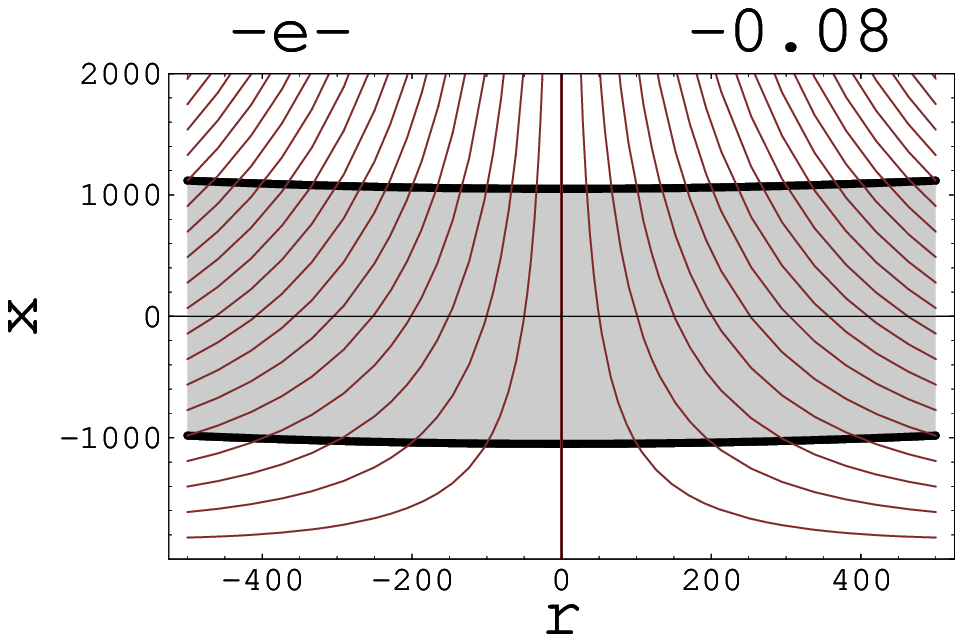}}
\caption{\footnotesize{Wavefronts distribution for different values of $\xi =
  0.0,0.04,0.08,-0.04,-0.08$. For a clearer visualization the variable
$r = 10^{2} \eta$ has been used. The shady region corresponds to the
  strip $|x| < \frac{1}{8 \beta}$ for the figure a) and $\left|P(\xi)
  x + Q(\xi) \eta^{2} \right| < \frac{1}{8 \beta}$} for the figure b).}
\label{fig_dkpfront}
\end{figure}

The three-dimensional generalization of the previous case has been
discussed in the same paper \cite{Solutions}. In particular, from the
formula (35)
one gets the following real phase function at $\lambda = 0$
\begin{equation}
\label{phase_3D}
\left. \hat{S}(x,\eta,\xi) \right|_{\lambda=0} = 2 \eta \left(P(\xi) x + Q(\xi) \eta^{2}\right)
\end{equation}
where
\begin{eqnarray}
P(\xi) &=& \frac{|q_{0}| \left(2 p_{0} - q_{0}^{2}
  \right)}{q_{0}^{2} \left(q_{0}^{2} - 4 p_{0}
  \right)^{\frac{1}{2}}} - 1, \nonumber \\
Q(\xi) &=& \frac{8 p_{0}}{3 q_{0}^{2}} \left(\frac{p_{0} -
  q_{0}^{2}}{\left(q_{0}^{2} -4 p_{0} \right)^{\frac{1}{2}}}-
  \frac{q_{0}^{2} \left(q_{0}^{2}- 3 p_{0} \right)}{|q_{0}|
  \left(q_{0}^{2} - 4 p_{0}
  \right)^{\frac{3}{2}}} \right), \nonumber \\
p_{0} &=& 1 + 36 \xi  - \left(1 - 12 \xi
  \right)^{\frac{3}{2}},~~~~~q_{0} = 3 \left(1 + 4 \xi \right). \nonumber
\end{eqnarray}
Note that for $\xi = 0$ the expression for $\hat{S}$~(\ref{phase_3D})
coincides with expression ~(\ref{phase_2D}). 
Moreover,the corresponding dKP solution is
\begin{equation}
\label{dkp_index_3D}
q = \frac{4 \left(5 - 12 \xi + 4 \left(1-12 \xi \right)^{\frac{1}{2}}
  \right)^{2} \left( \left(-1+12 \xi \right) x - 4 \eta^{2} \right)}{3
  (1+ 4 \xi) \left(1-12 \xi + 2 \left(1 - 12 \xi \right)^{\frac{1}{2}} \right)^{3}}.
\end{equation}
Of course, expression~(\ref{dkp_index_3D}) becomes~(\ref{dkp_index_2D}) when $\xi = 0$.
We discuss the $\xi$-deformations of wavefronts in a sufficiently small
neighbourhood of the plane $\xi = 0$, in order to avoid
$\xi$-values for which $q$ blows up or becomes complex valued. 

The explicit expression of the wavefronts is the following
\begin{equation}
\label{dkp_wfront_3D}
x = - \frac{\eta + 2\beta Q(\xi) \eta^{2} - k}{2 \beta P(\xi) \eta}
\end{equation}
where $k$ is an arbitrary constant.
The wavefronts~(\ref{dkp_wfront_3D}) should be considered inside a
strip given by
\begin{equation}
\left|P(\xi) x + Q(\xi) \eta^{2} \right| \ll \frac{1}{2 \beta}.
\end{equation}
\begin{figure}
\includegraphics[width=4.5cm]{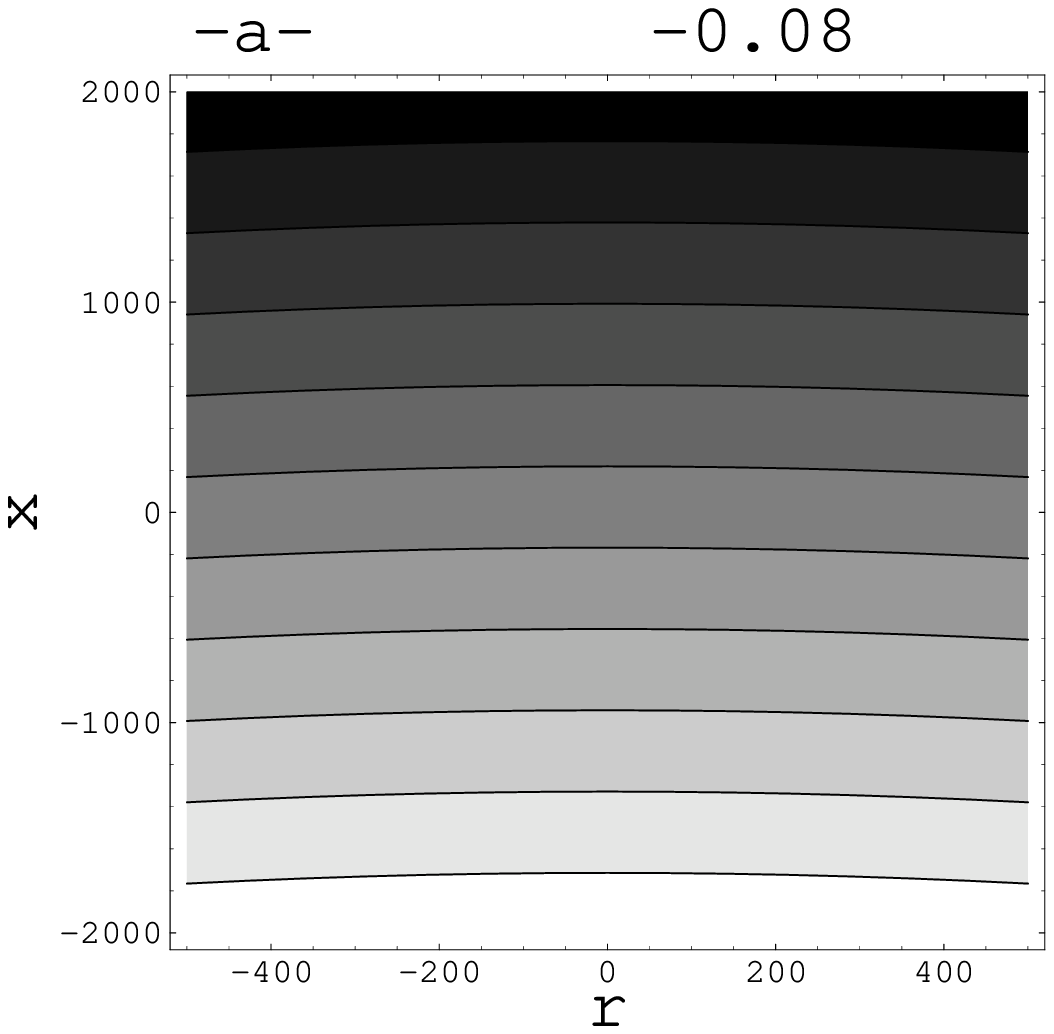}
\includegraphics[width=4.5cm]{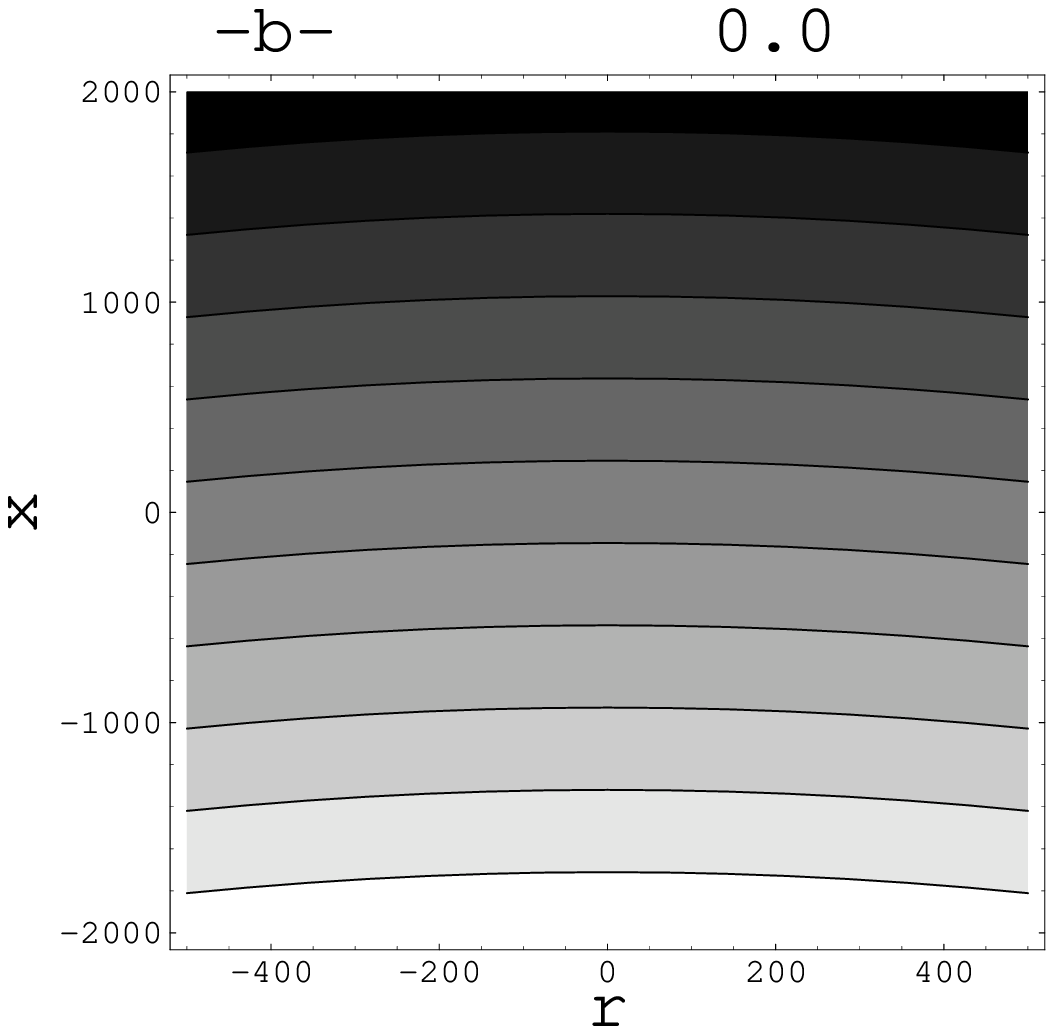}
\includegraphics[width=4.5cm]{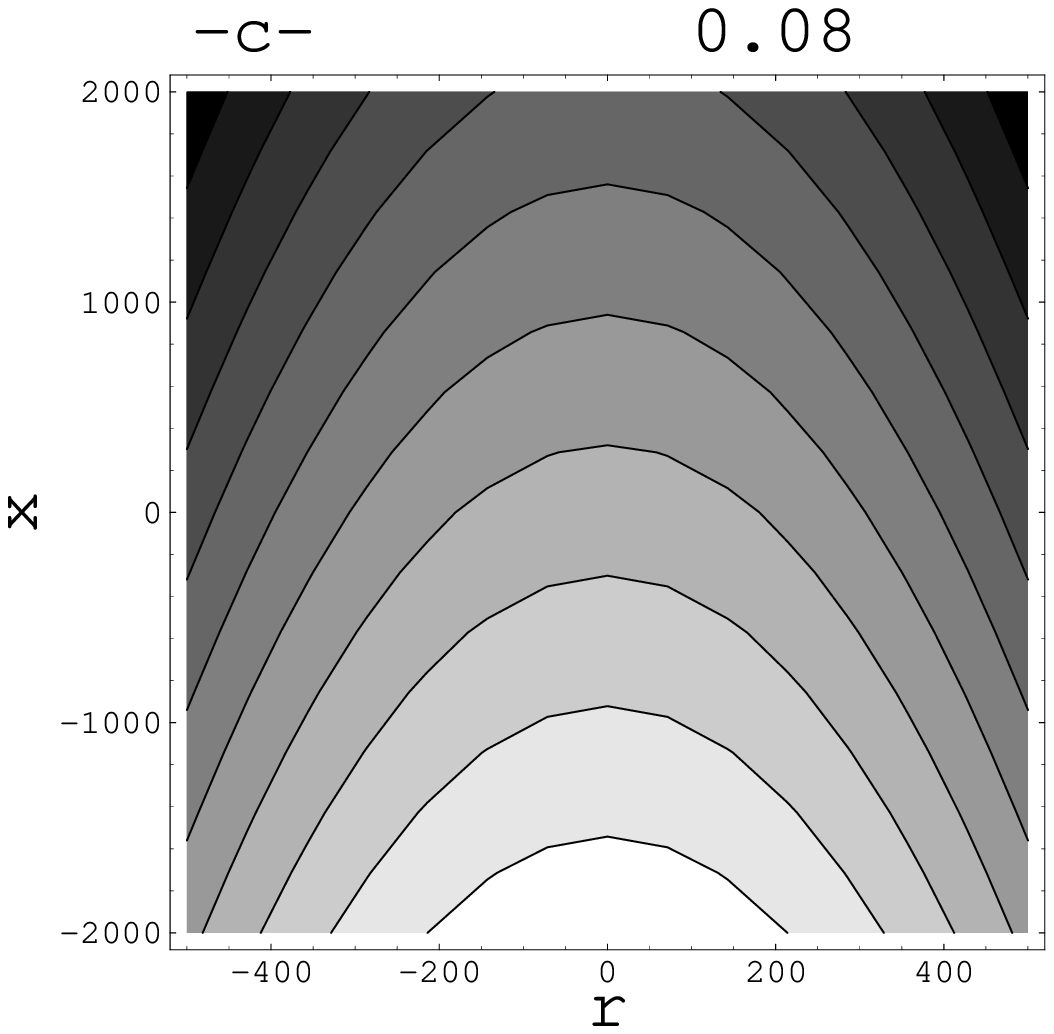}
\caption{\footnotesize{Density plot comparison of the refractive index $u$ for
    $\xi=-0.08,0.0,0.08$, where $r = 10^{2} \eta$. Darker regions correspond to a larger refractive index.} }
\label{fig_dkpindex}
\end{figure}
\begin{figure}
\centerline{\includegraphics[width=15cm]{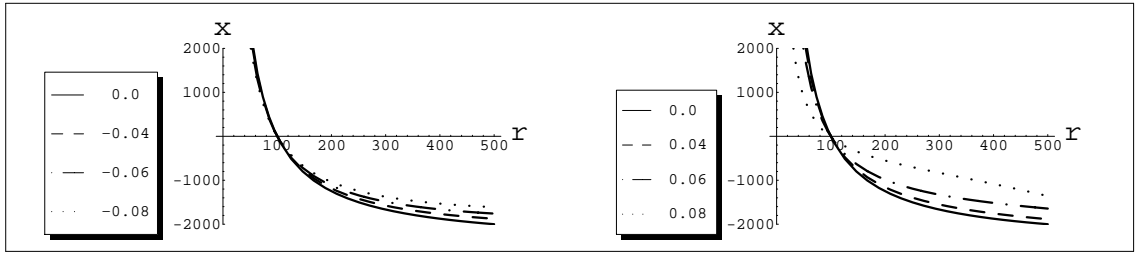}}
\caption{\footnotesize{Deformation of a wavefront at different values
    of $\xi$.}} 
\label{fig_dkpcomparison}
\end{figure}
\begin{figure}
\centerline{\includegraphics[width=8cm]{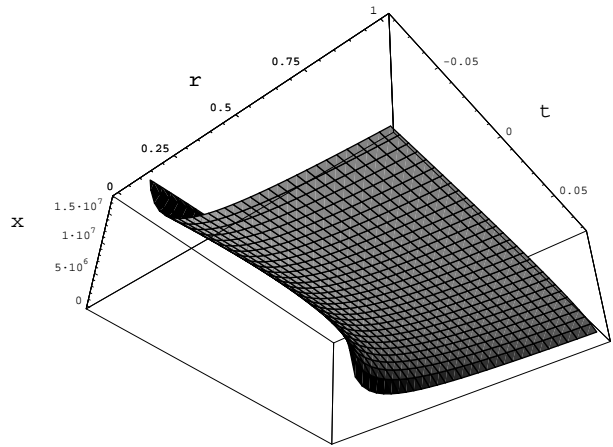}}
\caption{\footnotesize{Deformation of a wavefront along the $t \equiv
    \xi$ direction, where $r = 10^{2} \eta$. The wavefront bends in
    such a way that light rays turn
    off towards regions where the refractive index increases.}} 
\label{fig_dkpfrontdeform}
\end{figure}

The figure~\ref{fig_dkpfront} shows the wavefronts distribution inside a
reasonable strip for different $\xi$-values. The
figure~\ref{fig_dkpindex} demonstrates that the variation of refractive index is more
significant for $\xi > 0$, where we observe a larger deformation of
the wavefronts. One can see this in the figures~\ref{fig_dkpcomparison}
and~\ref{fig_dkpfrontdeform}.

\subsection{dVN wavefronts.}
Here, we analyze a refractive index
obtained by solution of the hodograph relation~(\ref{hodograph}), and a
corresponding numerical solution of eikonal equation. Note that
equation~(\ref{hodograph}) provides us with the complex refractive index  and,
by consequence,
the complex phase functions $S$, describing, in a way, damping effects of the
electromagnetic wave (see equation~(\ref{full_fields})). 
Assuming $u \neq 0$ and choosing, for simplicity, $\alpha^{3} = 3$, we
have the following form of equation~(\ref{hodograph})
\begin{equation}
\label{hodograph2}
-x + i y + (x + i y) u^{2} + 4 x_{2} u^{5} - u^{2} \psi(u) = 0. 
\end{equation}
A particular choice of the function $\psi(u)$ sets the behaviour of
$u$ on the boundary $x_{2} =0$. Assuming $\psi$ polynomial on $u$,
the numerical analysis shows 
qualitatively similar behaviours for different polynomial degrees.
As particular example, we set  $\psi = 2 u + \frac{1}{3} u^{2}$.
The figure~\ref{fig_dvnindex} shows the density plot of the real and
imaginary parts of a
refractive index $n = \frac{u}{4}$ on the transverse section $x_{2} =0$ , obtained by numerical solution
of~(\ref{hodograph2}).
\begin{figure}
\centerline{\includegraphics[width=5cm]{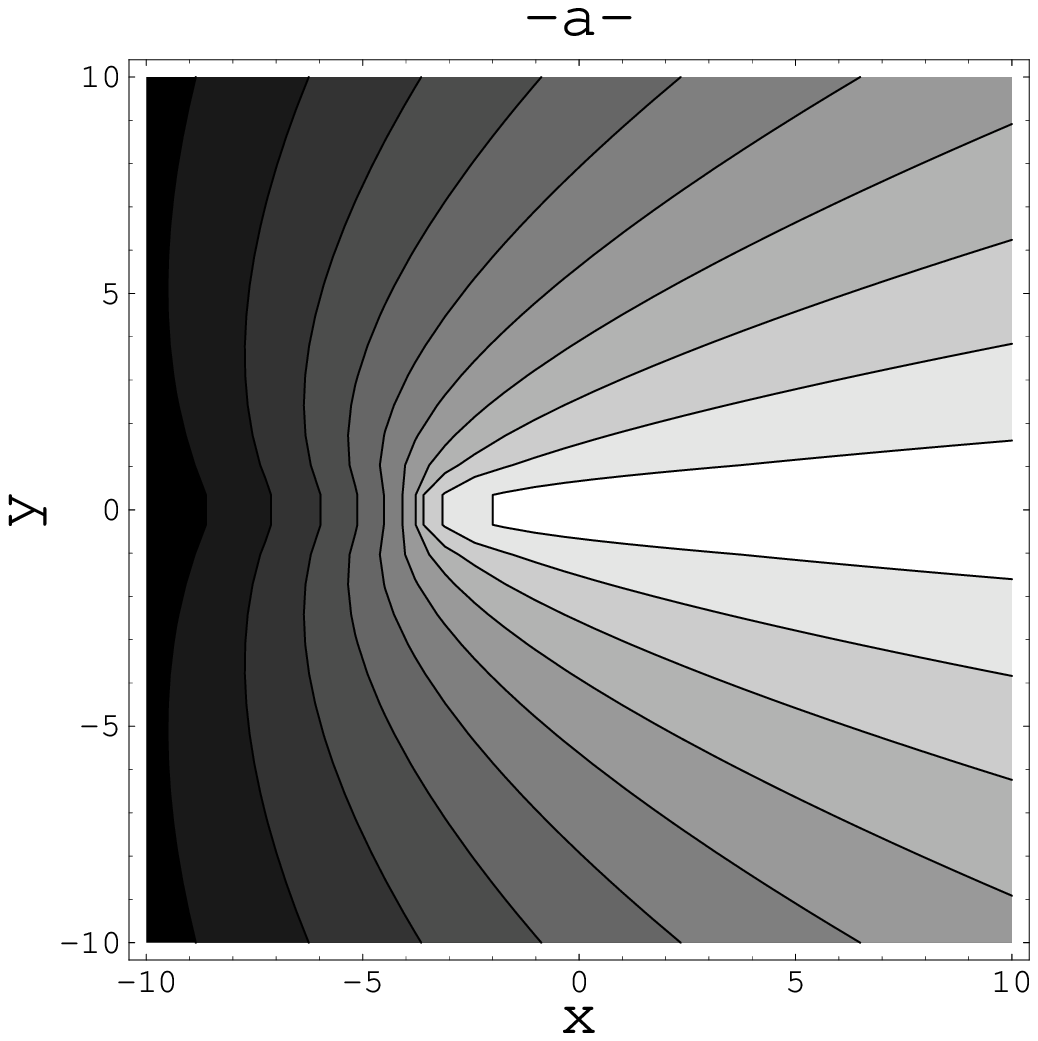}
\includegraphics[width=5cm]{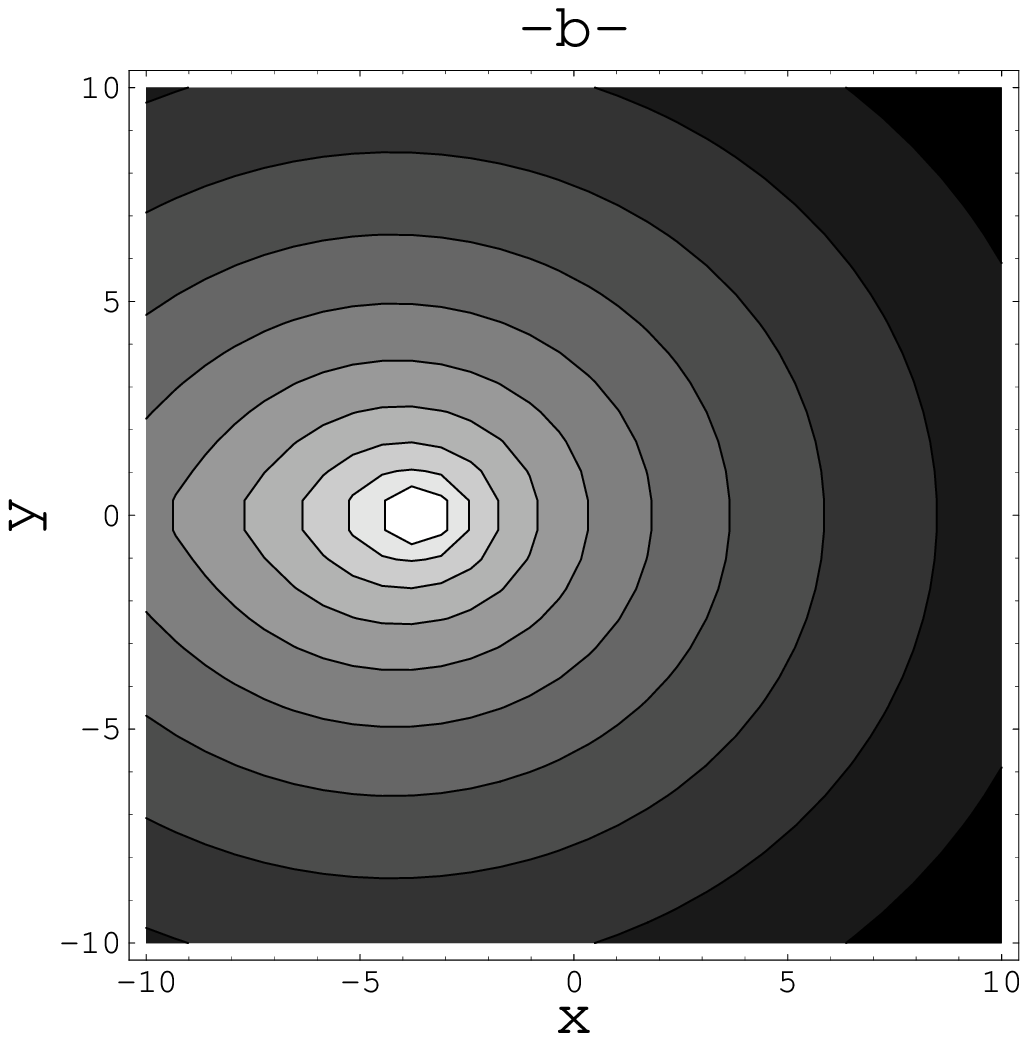}}
\caption{\footnotesize{Density plot of the a) real and b) imaginary
    parts of the refractive index $n$ on the plane $x_{2} = 0$.}}
\label{fig_dvnindex}
\end{figure}
\begin{figure}
\centerline{\includegraphics[width=6cm]{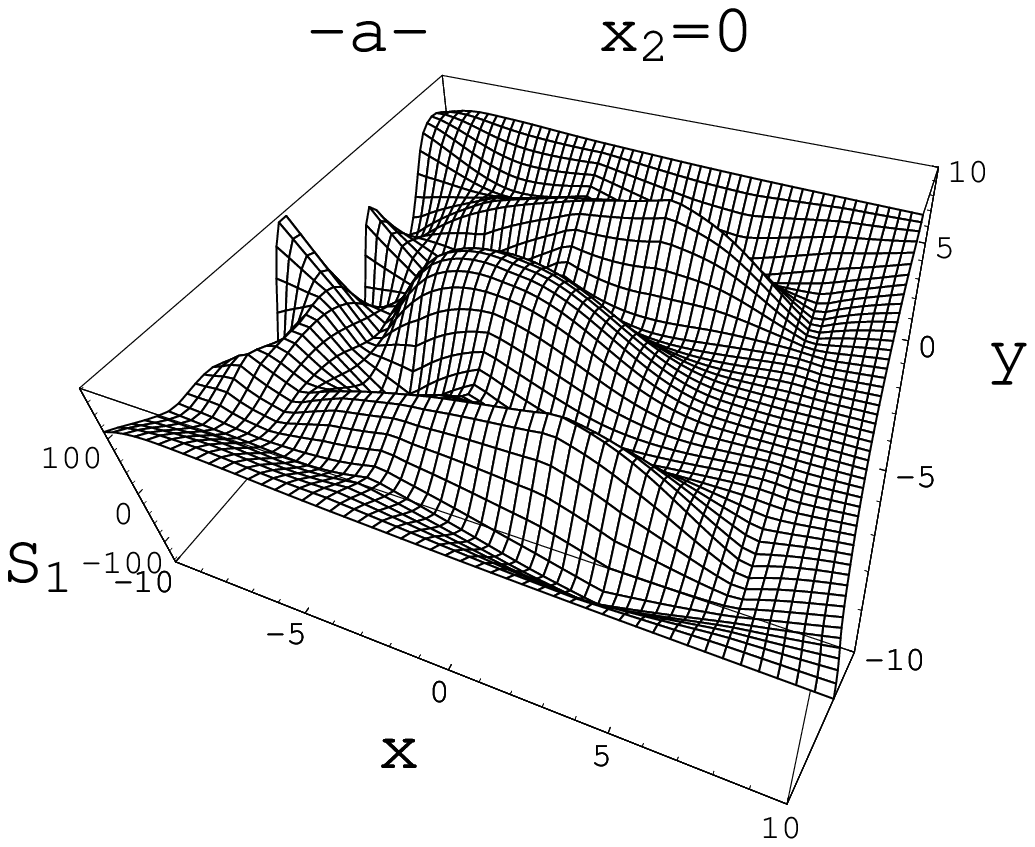}
\includegraphics[width=6cm]{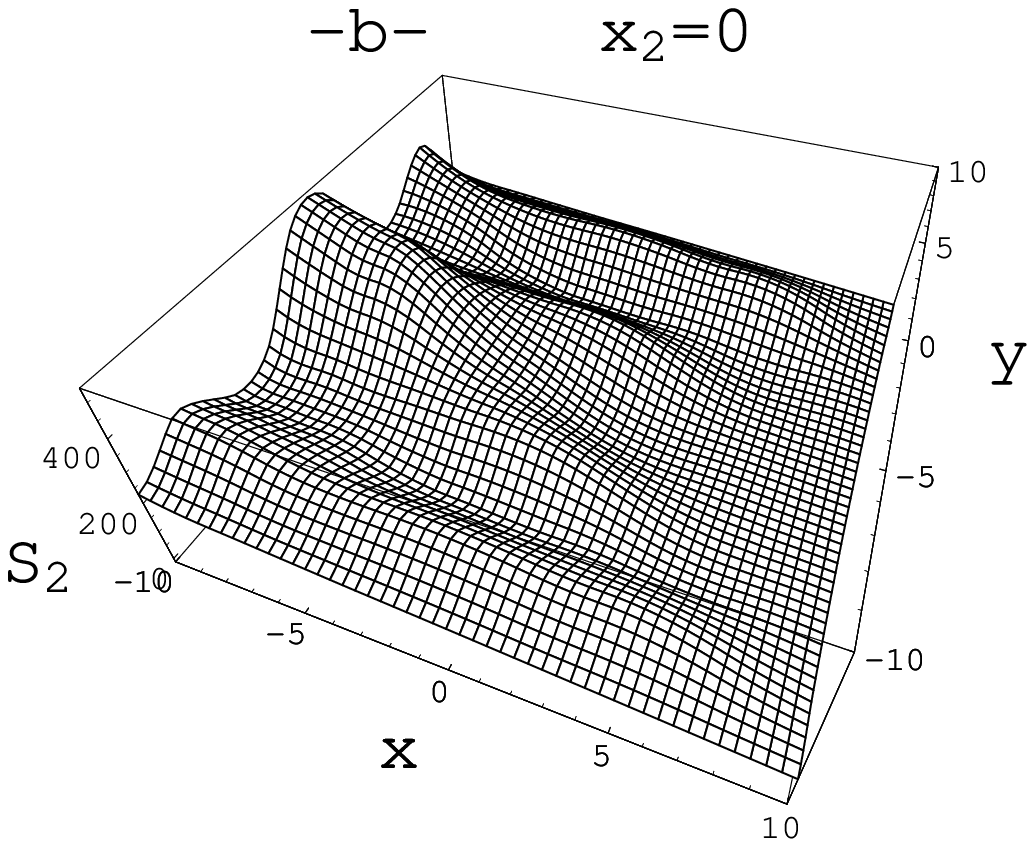}}
\caption{\footnotesize{a) Real part of the phase function with boundary conditions
  $S_{1}(10,y) = y^{2}$ and $S_{1}(x,-10) = S_{1}(x,10)$. b) Imaginary
  part of phase function with boundary conditions $S_{2}(10,y) = y^{2}$
  and $S_{2}(x,-10) = S_{2}(x,10)$.}} 
\label{fig_dvnsurface}
\end{figure}
\begin{figure}
\centerline{\includegraphics[width=5cm]{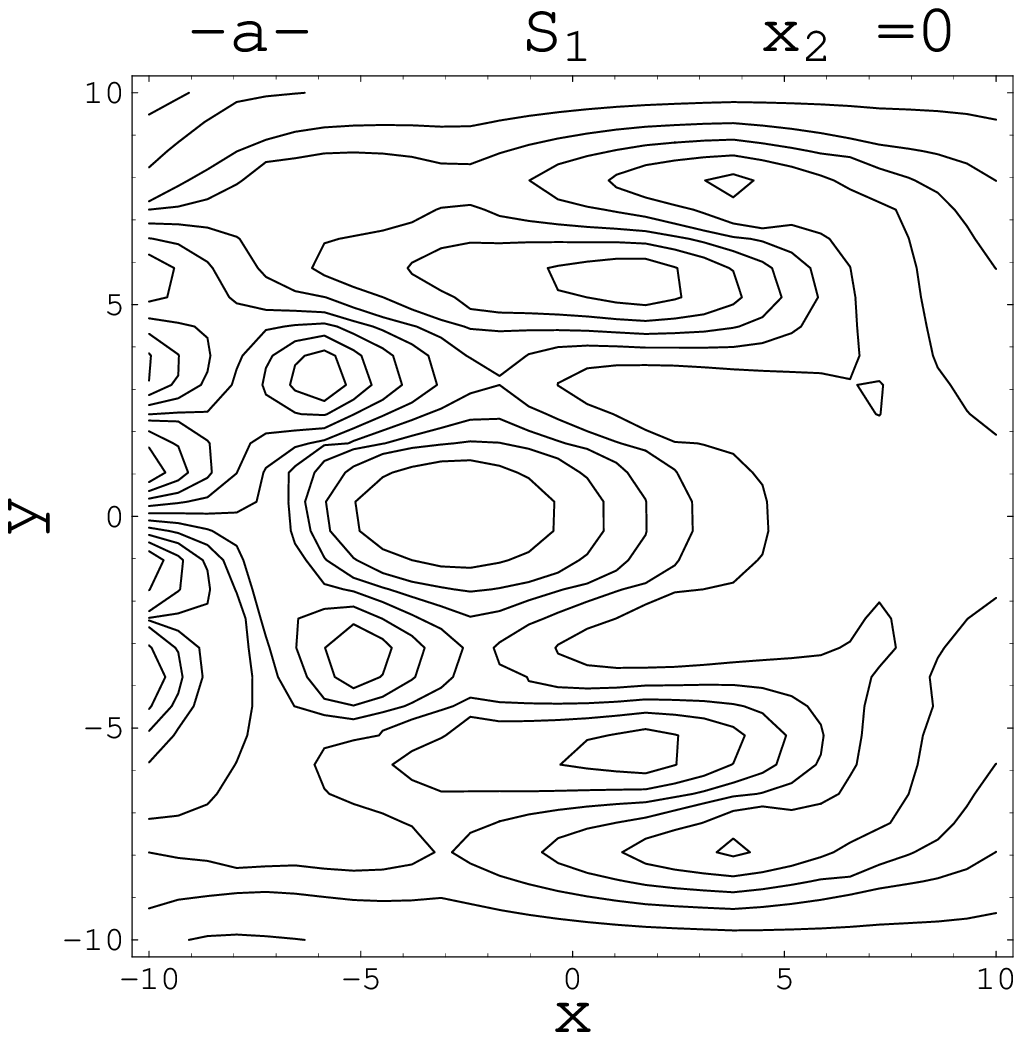}
\includegraphics[width=5cm]{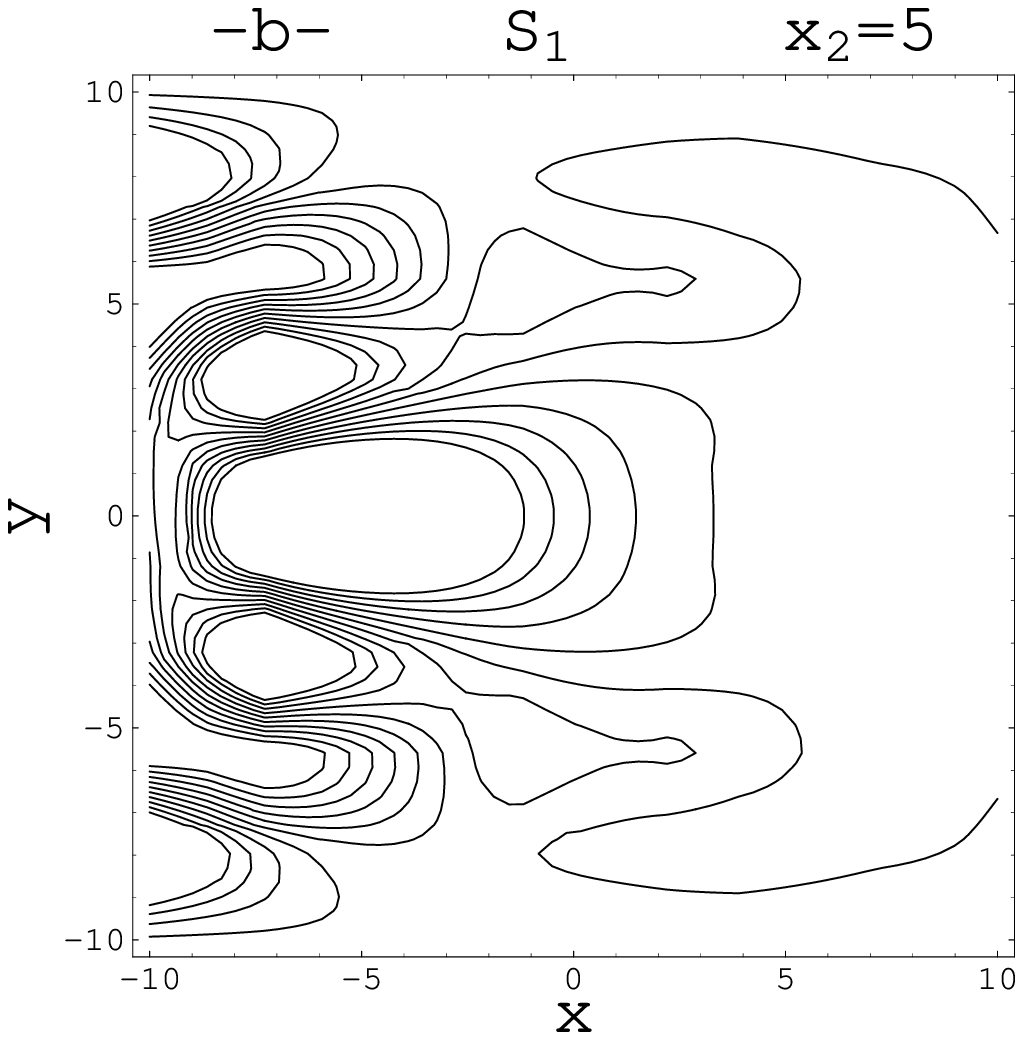}
\includegraphics[width=5cm]{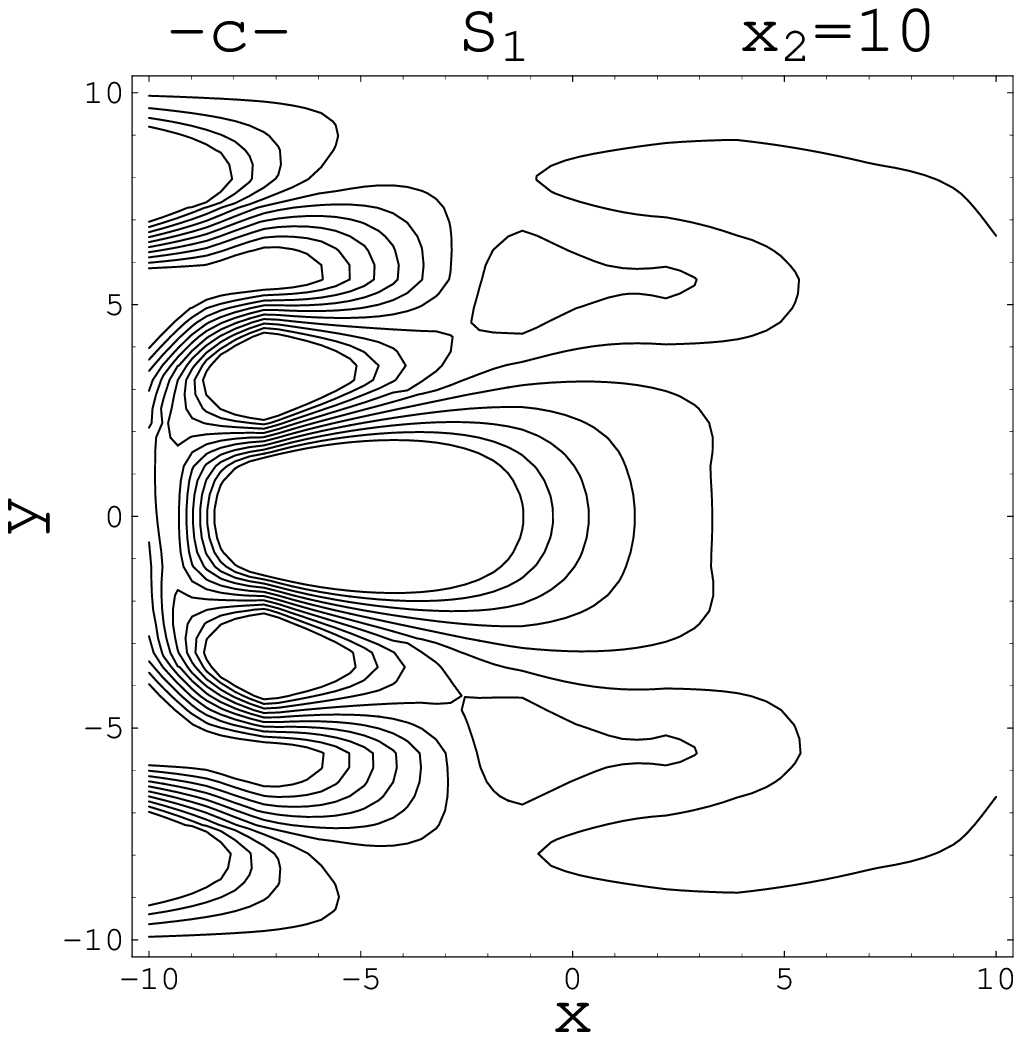}}
\caption{Wavefront configurations associated with the real phase $S_{1}$
  at different $x_{2}=0,5,10$.}
\label{fig_dvnfronts}
\end{figure}
Representing the phase function in
terms of its real and imaginary parts, $S=S_{1} + i S_{2}$, one rewrites the eikonal
equation~(\ref{system_a}) as the system
\begin{eqnarray}
\label{eikonal_cartesian}
&&S_{1 x}^{2} + S_{1 y}^{2} - S_{2 x}^{2} + S_{2 y}^{2} = 4 u_{1}\nonumber \\
&&S_{1 x} S_{2 x} + S_{1 y} S_{2 y} = 2 u_{2}
\end{eqnarray} 
where $u = u_{1} + i u_{2}$.
A numerical solution of the system~(\ref{eikonal_cartesian}) for the
refractive index plotted in the figure \ref{fig_dvnindex}. It has been
obtained in the range $x \in [-10,10]$ and $y \in [-10,10]$
exploiting the function {\tt NDSolve} of the {\tt Standard 
  Mathematica Packages}~\cite{Mathematica},
with boundary conditions $S_{1}(10,y)=S_{2}(10,y)=y^{2}$
and $S_{1}(x,-10) = S_{1}(x,10)$, $S_{2}(x,-10) = S_{2}(x,10)$. The
figures~\ref{fig_dvnsurface}-a) and ~\ref{fig_dvnsurface}-b) shows
$S_{1}$ and $S_{2}$ respectively. In particular, one should to note that
the regions where $S_{2}$ increases, correspond to the damping of the
electromagnetic wave.

Finally the figures~\ref{fig_dvnfronts}-a)-b) and c) show the
wavefronts configuration at different values $x_{2} = 0,5,10$. 

\bigskip

{\bf Acknowledgment.} The authors are grateful to L.V. Bogdanov for
the useful discussions.

\end{document}